\documentstyle[12pt,epsfig,longtable,amssymb]{article}
\setlongtables
\makeatletter
\@addtoreset{equation}{section}
\makeatother

\newcounter{appnum}
\def\theappnum{\Alph{appnum}}
\makeatletter
\def\makeappendix#1{%
\stepcounter{appnum}
\addcontentsline{toc}{section}{\appendixname\space\theappnum #1}
\setcounter{equation}{0}
\section*{\appendixname\ \theappnum #1\@mkboth
 {\appendixname\ \theappnum}{\appendixname\ \theappnum}}
\let\thesection\theappnum}
\def\appendixname{Appendix}
\makeatother
\begin{document}
\hspace*{10cm}PRA-HEP 99/8
\vspace*{1cm}
\begin{center}

{\large\bf High-energy elastic hadron collisions
and space structure of hadrons}\\[2cm]
Vojt\v{e}ch Kundr\'{a}t, Milo\v{s} Lokaj\'{\i}\v{c}ek\\[0.5cm]
{\it Institute of Physics,
AS~CR, 182 21 Praha 8, Czech Republic}\\[1cm]
Dalibor Krupa\\[0.5cm]
{\it Institute of Physics,
SAS, 84228 Bratislava, Slovakia}\\[5cm]

\begin{abstract}
The elastic and inelastic profiles have been derived
from the measured elastic differential cross section
data with the help of exact impact parameter
representation of elastic scattering amplitude. 
The results obtained for $pp$ scattering at energy of 53 GeV
and $\bar{p} p$ scattering at energy of 541 GeV have been 
presented. They indicate that nucleons can be regarded as
objects characterized by a small core (diameter $\sim$ 
0.4$\div$0.8 fm) and a half transparent outer shell 
(responsible mainly for elastic hadron scattering).
\end{abstract}
\end{center}
\newpage
\section{Introduction}

High-energy elastic hadron collisions are realized 
mainly at small absolute 
values of four-momentum transfer squared $|t|$. 
Corresponding differential cross sections exhibit a 
typical diffractive behavior characterized 
by approximate exponential decrease for
small $|t|$, being followed by diffractive minimum.
That has invoked a natural expectation that 
the hadron elastic processes are peripheral
(i.e., correspond to higher impact parameter $b$ values)
in contradistinction to inelastic processes
exhibiting a central character. 
However, the majority of models have described 
the high-energy elastic hadron collisions 
as central, i.e., a significant part 
of elastic hadron collisions should occur at 
$b \sim 0$ (see, e.g., Refs. \cite{miet}-\cite{kun2});
hadrons have been interpreted as transparent 
objects, which has represented a puzzle \cite{giac}.

The centrality of elastic processes has been
shown \cite{kun2} to be 
a direct consequence of one basic assumption
being involved in these models:
imaginary part $\Im F^N(s,t)$ of elastic 
hadron scattering amplitude being 
regarded as dominant 
in a broad interval of $t$ around zero
and vanishing only at the diffractive minimum.
However, the dominance of imaginary part can be 
justified theoretically in a very small
region of momentum transfers around forward direction only
\cite{kun1} at very high $\sqrt {s}$ 
center-of-momentum total energies. 
That does not eliminate
peripheral behavior requiring 
$\Im F^N(s,t)$ to go to zero approximately at
$|t| \sim 0.1$ GeV$^2$ (see  Ref. \cite{kun2}).

The standard description of elastic hadron scattering
involves, however, one puzzle more.
It concerns the fact that diffractive production
processes are being described as peripheral
(see Refs. \cite{miet,giov,chou}) even if they exhibit 
the characteristics very similar to elastic processes
(regarded at the same time as central) and differing significantly
(like elastic hadron collisions) from inelastic 
non-diffractive ones.

All these problems seem to follow from the fact
that only the modulus $|F^N(s,t)|$ can be determined
in the whole $t$ region from elastic experiments
almost uniquely while amplitude phase remains
arbitrary at least to some extent.
Its $t$ dependence can be estimated in principle
from a small part of elastic scattering 
data only, i.e., from a narrow
interval of small $|t|$ where Coulomb and hadron
scatterings interfere. Making use of 
sufficiently general parameterization of
the $t$ dependent elastic hadron amplitude
$F^N(s,t)$ the peripheral interpretation
of elastic hadron processes have been shown to be 
preferred \cite{kun2}, even if some characteristics 
cannot be determined quite uniquely.
Nevertheless, some new features of colliding
nucleons can be derived.

We will start by introducing corresponding
basic formulas enabling to relate the distributions
of elastic and inelastic processes in the
impact parameter space to the shape of 
elastic hadron scattering amplitude $F^N(s,t)$ 
(Sect. 2). Average values 
of the squares of impact parameter
values (mean-squares) for different
processes can be then shown to
be derived directly from the $t$ dependent elastic 
hadron scattering amplitude $F^N(s,t)$
in Sect. 3.

Sect. 4 describes the results of the analysis performed
on  $pp$ elastic scattering data at energy of 53 GeV and on
$\bar{p} p$ elastic scattering data at energy of 541 GeV.
It will be shown that unique values of the total cross section 
$\sigma_{tot}(s)$, the difractive slope $B(s)$ and 
the quantity $\rho(s,0)$, i.e., the ratio of the real
and imaginary parts of elastic hadron scattering amplitude
in forward direction cannot be derived. Only some 
admissible values can be established. In addition to, the 
root-mean-squared values of the impact parameters
characterizing the range of forces responsible
for the total, elastic and inelastic scatterings
will be derived, too. Some preliminary results
concerning $\bar{p} p$ scattering case have been 
presented already in Ref. \cite{kunc}.
The actual impact parameter 
profiles describing  the $\bar{p} p$ elastic 
hadron scattering at energy of 541 GeV as peripheral 
process in the impact parameter space
will be presented then in Sect. 5.

In Sect. 6 the problem of transparent nucleons in 
'head-on' collisions is discussed.
Transparent nucleons are to be considered as
an {\it{artifact}} having nothing to do with 
the physical reality; such a property follows
directly from the {\it{a priori}} assumption of
imaginary-part amplitude dominance. 

All presented results are based on
the rigorous representation
of elastic scattering amplitude in the impact parameter
space (valid at any $s$ and $t$)
proposed by Adachi and Kotani \cite{adac}
(our approach being based on), being described in Appendix A.  

\section{Elastic hadron scattering amplitude  
and impact parameter representation}

If the spins of colliding hadrons
are not taken into account the elastic hadron scattering
can be described fully with the complex
elastic hadron scattering amplitude $F^N(s,t)$. 
This amplitude is bounded together with
all production amplitudes $T(s,t, ...)$
by unitarity condition 
\begin{equation}
\Im <\!f|F^N|i\!> = \sum_{|e><e|} <\!f|F^N|e\!><\!e|F^N|i\!>  +
\sum_{|n><n|} <\!f|T|n\!><\!n|T|i\!>,
\label{ue1}
\end{equation}
where $\sum_{|e><e|}$ stands for an integration 
over all possible elastic intermediate states $|e\!\!>$
and  $\sum_{|n><n|}$ includes summation
over all possible production (inelastic) states  
as well as integration over all other kinematical variables;
$|i\!>$ and $|f\!>$ are initial and final states.
Eq. (\ref{ue1}) can be rewritten according to  van Hove (see 
Refs. \cite{hov2} or \cite{isla}) as
\begin{equation}
\Im F^{N}(s,t)= {p \over {4 \pi \sqrt{s}} }
\int d\Omega ' F^{N*}(s,t')F^{N}(s,t'') + G_{inel}(s,t),
\label{tcs1}
\end{equation}
where $G_{inel}(s,t)$ is the so called inelastic overlap function
and $p$ is the value of momentum of one particle  
in the center-of-momentum system;
$d\Omega' = \sin{\theta'}d\theta' d\Phi'$,
$t' = 2 p \sin {\theta'/2}$, $t''=2 p \sin {\theta''/2}$
and
$\cos{\theta''} = \cos{\theta} \cos{\theta'} +
\sin{\theta} \sin{\theta''} \cos{\Phi'}$.
According to the optical theorem the imaginary part of
the elastic hadron scattering amplitude in the forward direction
is related to the total cross section by
\begin{equation}
\sigma_{tot}(s) = {{4 \pi}\over {p \sqrt{s}}} \Im F^{N}(s,0).
\end{equation}

The elastic scattering amplitude $h_{el}(s,b)$
in the impact parameter space
can be obtained with the help of 
Fourier-Bessel (FB) transform 
of the elastic hadron scattering 
amplitude $F^{N}(s,t)$
\begin{equation}
h_{el}(s,b) = {1 \over {4 p \sqrt{s} }} 
\int\limits_{t_{min}}^{0}\! dt \;
J_0 (b \sqrt{-t}) \; F^{N}(s,t),
\label{hn1}
\end{equation}
where $J_0(x)$ is the Bessel function of the zeroth
order. At finite energies the region of kinematically allowed
$t$ values is limited: $t \in (t_{min}, 0)$; e.g., 
in the case of nucleon - nucleon scattering $t_{min}= -s+4m^2$,
where $m$ is the nucleon mass. In order to
define the FB transform exactly
(existence of inverse transform) 
at finite energies, the elastic hadron scattering 
amplitude is to be defined also in the unphysical
region of $t$, i.e., in the interval 
$(-\infty, t_{min})$.
Consequently, the actual shape of
elastic hadron scattering amplitude
$h_{el}(s,b)$ determined with the 
help of Eq. (\ref{hn1}) might be deformed
in an uncontrolled way. These problems
were solved by Adachi and Kotani \cite{adac}
and by Islam \cite{isla};
the corresponding mathematically exact theory 
of elastic scattering amplitude is
briefly described in the Appendix A. 

The amplitude $h_{el}(s,b)$ has been
devided into two terms. The first contribution 
$h_{1}(s,b)$ is determined by the FB transform 
of elastic hadron amplitude defined in the physical
region of $t$ while the second one comes from
the unphysical region. The FB transform
of the inelastic overlap function $G_{inel}(s,t)$
has been devided in a similar way.      
The unitarity condition (\ref{mv14}) in the 
impact parameter space contains only
the physically relevant terms based on
the FB transforms of the functions $F^N(s,t)$ and
$G_{inel}(s,t)$ in the physical region of $t$;
and also the correction term is defined with
the help of amplitude values from the physical
region.

The elastic profile $|h_{1}(s,b)|^2$ is always 
non-negative and generates the total elastic cross section;
it can be, therefore, considered as the 
distribution of elastic collisions at the impact parameter $b$.
The other profiles, i.e., $\Im h_{el}(s,b)$ and
the FB transform of $G_{inel}(s,t)$ 
oscillate around zero for higher impact parameter values.
They cannot be, therefore, considered as the corresponding
distributions even if their
integrals performed over all impact 
parameter values give the corresponding cross sections.

However, as shown in the Appendix A, physically 
relevant profiles can be derived.
Using the ambiguity of the FB transform
of inelastic overlap function $G_{inel}(s,t)$ in the unphysical
region of $t$, the unitarity condition 
(\ref{mv14}) can be modified
in such a way that the new total and 
inelastic profiles can be made to be non-negative, too.
The new unitarity condition (\ref{mv15}) is obtained
where all individual profiles are 
represented by the non-negative distribution 
functions in the impact parameter space.

As already mentioned in Sec. 1 the 
currently used models of elastic scattering 
have led to the central impact parameter 
distribution of Gaussian type 
with the center at $b = 0$ (see, e.g., Refs.
\cite{miet}-\cite{kun2}); the distribution 
being narrower than that exhibited
by inelastic processes (see, e.g., Ref. \cite{miet}).
Such a picture has followed directly from 
the {\it{a priori}} assumption requiring the real
part of amplitude to be neglected in a 
broad interval of $t$ around zero. 
If such a limitation is not applied to
better fits of experimental data
can be obtained. In such a case
the elastic hadron scattering should be interpreted
as a peripheral process.
However, even if the peripherality is to
be preferred the experimental data
may hardly lead to unique picture, as
equivalent $\chi^2$ values can be obtained with
different degrees of peripherality;
the peripherality degree being characterized, 
e.g., by the ratio of
root-mean-squares of impact parameters
for elastic and inelastic processes.
Even if the available experimental data
do not allow to establish these values 
quite uniquely, they represent surely a more
qualified physical characterization
of hadron collision processes than
the values of $B$ and $\rho$ (being added to
$\sigma_{tot}$).

\section{Mean-squares of impact parameter distributions}
The mentioned mean-square of elastic impact parameter 
distribution can be defined as
\begin{equation}
<b^2(s)>_{el}\;\; = \;\;{{\int \limits_{0}^{\infty} b db\; b^2 \;
|h_{1}(s,b)|^2}
\over {\int \limits_{0}^{\infty}b db \; |h_{1}(s,b)|^2}},
\label{mv1} 
\end{equation}
where $h_1(s,b)$ is the FB transform of elastic 
hadron scattering amplitude $F^{N}(s,t)$ in the
physical region of $t$ (see Appendix A, 
Eqs. (\ref{el1}) - (\ref{el2})).
This quantity can be expressed in a form containing
the $t$ dependent elastic hadron scattering amplitude 
\cite{kun2,hene}:
\begin{equation}
<b^2(s)>_{el}\;\; = \;\;4 \;\;{{\int \limits_{t_{min}}^{0}\!\! dt \; |t| \;
|{d \over{dt}} F^{N}(s,t)|^2}
\over { \int \limits_{t_{min}}^{0}\!\!dt \; |F^{N}(s,t)|^2}},
\label{mv2}
\end{equation}
which can be rewritten further as a sum of the two terms
\begin{eqnarray}
\!\!\!\!\!\!\!\!\!\!\!<b^2(s)>_{el}  \; &=&  \;
 <b^2(s)>_{mod} + <b^2(s)>_{ph}  =
\nonumber   \\ 
&=& 4 \; {{\int \limits_{t_{min}}^{0}\!\! dt \;|t| \;
\Bigl({d \over{dt}} |F^{N}(s,t)|\Bigr)^2}
\over { \int \limits_{t_{min}}^{0}\!\! dt \; |F^{N}(s,t)|^2}} +
4 \; {{\int \limits_{t_{min}}^{0}\!\! dt \;|t| \;
|F^{N}(s,t)|^2 \Bigl( {d \over {dt}} \zeta^{N}(s,t)\Bigr)^2} 
\over { \int \limits_{t_{min}}^{0} \!\!dt \; |F^{N}(s,t)|^2}},
\label{mv3}
\end{eqnarray}
where the contributions of the modulus $|F^{N}(s,t)|$
and the phase $\zeta^{N}(s,t)$ of the
elastic hadron amplitude $F^{N}(s,t)$ defined by
\begin{equation}
F^{N}(s,t) = i |F^{N}(s,t)| e^ {-i \zeta^{N} (s,t)},
\label{nu1}
\end{equation}
are separated and both are non-negative. The derivation of
Eq. (\ref{mv2}) given in Ref. \cite{kun2} enabled its
generalization to the case when the spins of all
particles involved are taken into account \cite{com2}.

The mean-square of the total impact parameter
(i.e., for all collision processes) 
can be defined in analogy to Eq. (\ref{mv1}) as
\begin{equation}
<b^2(s)>_{tot}\;\; = \;\;{{\int \limits_{0}^{\infty} b db\; 
b^2 \; h_{tot}(s,b)}
\over {\int \limits_{0}^{\infty}b db \;  h_{tot}(s,b)}},
\label{mv4}
\end{equation}
where $h_{tot}(s,b)$ is the distribution of 
all collisions in the impact parameter space
- see Appendix, Eq. (\ref{mv15}).

Using the impact parameter representation of
the elastic hadron scattering amplitude introduced by
Adachi and Kotani one can write
\begin{equation}
{d \over {dt}} \ln \Im F^{N}(s,t) = 
{1 \over {2 \sqrt {-t}}}
\;\;{{\int \limits_{0}^{\infty} b db\; 
b^2 \; h_{tot}(s,b) \; J_{1}(b \sqrt{-t})}
\over {\int \limits_{0}^{\infty}b db \;  h_{tot}(s,b)} \; J_{0}(b \sqrt{-t})}, 
\label{mv5}
\end{equation}
where $J_{1}(x)$ is the Bessel function of the first order. 
Going to limit $\sqrt {-t} \rightarrow 0$ in Eq. (\ref{mv5}) 
one obtains
\begin{equation} 
{{d} \over {dt}} \ln \Im F^{N}(s,t) \bigg {|}_{t=0} =
{ 1 \over 4}<b^2(s)>_{tot}.
\label{mv6}
\end{equation}
On the other hand the logarithmic derivative of
$\Im F^{N}(s,t)$ can be also expressed with the 
help of its modulus and phase (see Eq. (\ref{nu1})) as
\begin{equation}
{{d} \over {dt}} \ln \Im F^{N}(s,t) = 
{ {{d \over {dt}} |F^{N}(s,t)|} \over {|F^{N}(s,t)|}} -
\cot \zeta^{N}(s,t) \; {d \over{dt}} \zeta^{N}(s,t).
\label{mv7}
\end{equation}
Assuming the first derivative of the phase
$\zeta^{N}(s,t)$ to vanish at $t=0$ (which is quite
plausible) we obtain from
Eqs. (\ref{mv5}) - (\ref{mv7}), that
(as generally $\zeta^{N}(s,0) \ne 0$)
\begin{equation}
<b^2(s)>_{tot}\;\; = 2 B(s,0),
\label{mv8}
\end{equation}
where $B(s,0)$ is the diffractive slope 
at $t=0$, defined generally as
\begin{equation}
B(s,t)= {2\over |F^{N}(s,t)|}\;{d\over {dt}}|F^{N}(s,t)|.
\label{sl1}
\end{equation}
Eq. (\ref{mv8}) was derived already in Ref. \cite{ida}
under two assumptions:
differential cross section having a purely
exponential $t$ dependence for small $|t|$ values,
and the real part of $F^N(s,t)$ being neglected.
Our derivation shows that it holds under
more general conditions.

The mean-squares of impact parameter for 
elastic and total processes have been 
determined by Eqs. (\ref{mv2}) and (\ref{mv8}).
The mean-square of inelastic impact parameter
can be then established with the help of modified unitarity 
equation (\ref{mv15}). 
Multiplying this equation by $b^3$ and integrating 
over all possible impact parameter values $b$
one obtaines 
\begin{equation}
\int \limits_{0}^{\infty}b db\; b^2 \; h_{tot}(s,b) = 
 \int \limits_{0}^{\infty}b db\; b^2 \; |h_{1}(s,b)|^2  \; + \;
 \int \limits_{0}^{\infty}b db\; b^2 \; g_{inel}(s,b).
\label{mv11}
\end{equation}
Defining the mean-square of the inelastic impact
parameter in analogy with Eqs. (\ref{mv1}) and (\ref{mv4}) as
\begin{equation}
<b^2(s)>_{inel}\;\; = \;\;{{\int \limits_{0}^{\infty} b db\; 
b^2 g_{inel}(s,b)}
\over {\int \limits_{0}^{\infty}b db \; g_{inel}(s,b)}}, 
\label{mv12}
\end{equation}
Eq. (\ref{mv11}) can be rewritten as 
\begin{equation}
<b^2(s)>_{tot}\;\; = \; 
{{\sigma_{el}(s)} \over {\sigma_{tot}(s)}} <b^2(s)>_{el} \; + \;
{{\sigma_{inel}(s)} \over {\sigma_{tot}(s)}} <b^2(s)>_{inel}.
\label{mv13}
\end{equation}
As both the elastic and total mean-squares can be
calculated directly from elastic amplitude $F^N(s,t)$
in $t$ variable with the help of 
Eqs. (\ref{mv2}), (\ref{mv8}) and (\ref{sl1}), Eq. (\ref{mv13})
can be used for establishing their inelastic analogue.
Eq. (\ref{mv13}) represents a rigorous result: the mean-square
of total impact parameter can be expressed as the weighted
sum of mean-squares of elastic and inelastic impact parameters
with the weights representing the corresponding branching ratios.

It is useful to mention that Eqs. (\ref{mv11}) and (\ref{mv13})
are valid not only for the second power of $b$ but
for any function of $b$ provided the corresponding
integrals exist. 

\section{Elastic nucleon - nucleon scattering 
and analysis of experimental data}

In the experiments with charged nucleons the hadron
scattering is always accompanied by
Coulomb scattering plying a dominant role at
very small $|t|$. The analysis of corresponding
experimental data (i.e., in the interference region)
was being performed in the past 
currently with the help of a simplified 
West and Yennie (WY) formula 
(see, e.g., Refs. \cite{west} - \cite{matt})
for the total amplitude $F^{C+N}(s,t)$
\begin{eqnarray}
F^{C+N}(s,t) &=&
e^{i\alpha \Psi} F^{C}(s,t) + F^{N}(s,t)
\nonumber \\
&=& \pm {\alpha s \over t} f_1(t)f_2(t)e^{i\alpha \Psi} +
{\sigma_{tot} \over {4\pi}} p\sqrt {s} (\rho+i)e^{Bt/2},
\label{wy1}
\end{eqnarray}
derived within the framework of QED with the
help of Feynman diagrams representing one photon exchange.
There are two form factors $f_1(t)$ and $f_2(t)$ corresponding
to individual colliding hadrons in the first term 
representing Coulomb elastic scattering amplitude
$F^C(s,t)$. The upper (lower) sign corresponds to the $pp$
$(\bar{p} p)$ scatterings. 
It holds \cite{west}
for the relative phase in the
standard WY expression (\ref{wy1}):
\begin{equation}
 \alpha \Psi = \mp \alpha (\ln (-Bt/2) + \gamma )
 \label{wy2}
\end{equation}
where $\gamma =0.577215$ is the Euler constant
and $\alpha = 1/137.036$ is the fine structure constant. 
The elastic hadron scattering
is fully characterized in this scheme
by the total cross section
$\sigma_{tot}$, the quantity $\rho$ 
(the ratio of the real to imaginary parts
of elastic hadron scattering amplitude in forward
direction) and the 
diffractive slope $B$ (all these quantities 
being constant).

There are, however, two main defficiencies 
in using Eq. (\ref{wy1})
for the analysis of experimental data.
First, different formulas must be made use of
for the lower and higher values of $|t|$
$(F^{C+N}(s,t)$ and/or $F^N(s,t)$),
while the boundary is not clearly defined.
Second, the formula is based on
some simplifying assumptions that are not
fully justified \cite{kun1}.

To obtain more precise and 
sufficiently consistent results
it is necessary to use a more general
formula for the total elastic scattering 
amplitude \cite{kun1,kun3}: 
\begin{eqnarray}
F^{C+N}(s,t)&=&\pm {\alpha s\over t}f_1(t)f_2(t) +
F^{N}(s,t)\;\bigg\{1\mp i\alpha \int\limits_{t_{min}}^0 \!\!
dt'\bigg[\ln{t'\over t}\;\bigg[f_1(t')f_2(t')\bigg]'
\nonumber \\
&& -{1\over {2\pi}}\bigg [{F^{N}(s,t')\over F^{N}(s,t)}-1\bigg] \;
I(t,t')\bigg]
\bigg\} ,
\label{ta1}
\end{eqnarray}
where
\begin{equation}
I(t,t')=\int\limits_0^{2\pi}d\Phi \;
{f_1(t^{\prime \prime})f_2(t^{\prime \prime})\over t^{\prime \prime}}
\label{ta2}
\end{equation}
and $t^{\prime \prime}=t+t'+2\sqrt{tt'}\cos\Phi$.
This formula is valid at any $s$ and $t$ up
to the terms linear in $\alpha$.
At difference to the standard analysis 
based on the simplified WY formula 
\cite{bloc,matt} the new approach enables to perform
the analysis in the whole region of 
experimental differential cross section data,
i.e., in the Coulomb, interference and hadronic domains
simultaneously with the help of one common formula.
It enables to separate the Coulomb and hadron
elastic scatterings practically in a model-independent way.
It turns out that contrary to general belief
the influence of Coulomb scattering even at
higher $|t|$ values cannot be fully neglected \cite{kun3}.

The corresponding differential cross section is given by 
\begin{equation}
{{d \sigma (s,t)} \over {dt}}= {\pi\over {sp^2}}|F^{C+N}(s,t)|^2.
\label{ds1}
\end{equation}

The total cross section $\sigma_{tot}$ equals now \cite{kun3}
\begin{equation}
\sigma_{tot}={{4\pi}\over {p\sqrt s}}|F^{N}(s,0)|
{1\over {\sqrt{1+\rho^2(s,0)}}},
\label{st1}
\end{equation}
where
\begin{equation}
\rho(s,t)= \tan \zeta^{N}(s,t)
\label{ro1}
\end{equation}
is the ratio of real to imaginary parts at individual 
values of $t$; the value of $\rho (s,0)$  
being compared to $\rho$ in Eq. (\ref{wy1}); and 
similarly $B(s,0)$ (see Eq. (\ref{sl1})) to $B$.

To derive the $t$ dependence of the elastic
hadron scattering amplitude from experimental
data (in the absence of any reliable theory of
diffraction scattering)
some convenient parametrization of its 
has to be used. The following parameterizations 
of modulus and phase (proposed in Ref. \cite{kun3})
has been made use of: 
\begin{equation}
|F^{N}(s,t)|=(a_1+a_2t)\exp (b_1t+b_2t^2+b_3t^3) +
(c_1+c_2t)\exp(d_1t+d_2t^2+d_3t^3),
\label{mo1}
\end{equation}
\begin{equation}
\zeta^{N} (s,t) = \zeta_0 + \zeta_1 {\bigg| {t\over t_0}\bigg| } ^{\kappa}
e^{\nu t} + \zeta_2 {\bigg| {t\over t_0}\bigg| } ^{\lambda},
\hspace{1cm} t_0 = 1 \;  GeV^2.
\label{ph1}
\end{equation}
Any behavior including central as well as 
peripheral pictures of elastic hadron scattering
may be then described with the help of these
formulas; the quantities $a_k, c_k, b_j, d_j,
\zeta_i, \kappa, \nu$ and $\lambda$ being free 
(energy dependent) parameters.

Some analysis has been performed also with the help of 
the phase parameterized as
\begin{equation}
\zeta^{N} (s,t) = \arctan {\rho_0  \over {1 - {\bigg| {{t\over t_{di\!f\!\!f}}
\bigg |}} }},
\label{ph2}
\end{equation}
where $t_{di\!f\!\!f}$ corresponds to diffractive minimum
($\rho_{0}$ and $t_{di\!f\!\!f}$ being 
energy dependent free parameters).
This formula always leads to a central distribution
of elastic hadron scattering.

It was already mentioned that the $t$ dependence
of the phase cannot be uniquely derived from
experimental data with the help of 
general parameterization (\ref{ph1}).
Consequently we have added the constraint
by requiring for the ratio
\begin{equation}
\eta = {{|h_{el}(s,b=0)|^2} \over{|h_{el}(s,b_{max})|^2}}
\label{eta}
\end{equation}
to have in the optimization procedure
a fixed value lying between 0 and 1.
Here, $b_{max}$ is the value of impact parameter
for which $|h_{el}(s,b)|^2$ has its maximum.
Evidently, the case when $b_{max} = 0$ (i.e., $\eta = 1$)
corresponds to a central picture, while $b_{max} > 0$
leads to a peripheral picture. The degree of peripherality will be
maximal when $\eta$ approaches 0. 

The given formulas have been applied to experimental data
on $pp$ elastic scattering at energy of 53 GeV and
$\bar{p} p$ elastic scattering at energy of 541 GeV.
Fits for different values of $\eta$ have been found.

\subsection{$pp$ elastic scattering at energy of 53 GeV}

The standardly normalized
data for $pp$ elastic scattering at
energy of 53 GeV were taken from Ref. \cite{byst}.
To combine the data from different experiments
the normalization coefficients in individually measured 
kinematical regions were considered as free
parameters in the fits. Their values were
admitted to change within
the corresponding statistical errors;
once established they 
were kept fixed in all remaining
constraint fits.

Table I contains the results of the analysis performed 
with the help of formula (\ref{ta1}) for 
both the cases of central and peripheral pictures of 
elastic hadron scattering in the impact parameter space.
The modulus of elastic hadron amplitude was parameterized
with the help of formula (\ref{mo1}) in 
both the cases; for its phase
formula (\ref{ph1}) was used in the peripheral case
while in the central case formula (\ref{ph2}) 
was applied to.

Using formula (\ref{ph2}) the phase $\zeta^N (s,t)$
can be uniquely established
from experimental data, and consequently,
the values of $\sigma_{tot}$, $B$ and $\rho$ 
can be uniquely determined from the available
experimental data. 
With the help of Eq. (\ref{ph1}) 
only some admissible regions of these values 
can be derived.

Fixing the quantity $\rho=\rho(s,0)$ to different values
we looked then for the best fits
in both the peripheral and central cases
of elastic hadron scattering.
The corresponding values of $\chi ^2$
(Table I)
indicate  that the admissible
values of $\rho$ in the peripheral case of $pp$  elastic
collisisons can lie in the interval (-0.12, 0.08) while  in the
central case the interval of admissible $\rho$ values
is much narrower, i.e., (0.07, 0.08) - see Fig. 1. 

Different values
of $\sigma_{tot}$ and $B$ correspond, of course,
to changing values of $\rho$. The values of
$\sigma_{tot}$ in the peripheral case
lie within the interval
(42.65, 42.97) mb  (see Table I and Fig. 2),
while the values of the diffractive slope
$B$ can lie in the interval 
(13.45, 13.68) GeV$^2$ (see 
Table I and Fig. 3). 
In the central case $\sigma_{tot}$
lies within  the interval (42.65, 42.75) mb 
(see Fig. 2) and $B$ in the interval (13.25, 13.35) 
GeV$^2$ (see Fig. 3).
The typical errors
corresponding to individual basic
quantities, i.e., to the total cross section,
the diffractive slope  and the $\rho$ quantity
are given in Table I, too
(quantities in brackets).

All the optimal peripheral fits (corresponding
to different values of $\rho$) exhibit 
roughly the same peripheral profiles
characterized by the value of $\eta \sim 0.38$ 
(see Eq. (\ref{eta}))
and also approximately by the same value of elastic 
root-mean-square $\sim 1.80$ fm
as shown in Table II and Fig. 4 
while in the central case the elastic 
root-mean-square is practically constant and
equals approximatelly only to 0.68 fm (see Fig. 5).

The value of the root-mean-square
of elastic impact parameter 
(1.78 $\div$ 1.80) fm calculated with the help of 
Eq. (\ref{mv3}) is composed of two terms.
The first term representing the contribution
of the modulus of the elastic hadron scattering amplitude
contributes to the elastic root-mean-square
by the value of 0.68 fm. 
The second term, containing a contribution of the 
phase of elastic hadron amplitude, represents then 
the main contribution to the root-mean-square
of elastic impact parameter in a peripheral case,
reaching the value of about 1.65 fm.
It equals practically zero for 
the central behavior.

Table II also contains the root-mean-square
values of the total
and inelastic impact parameters. The root-mean-squares
of the total impact parameter determined with 
the help of Eq. (\ref{mv8}) are slightly higher than 1 fm.
They do not depend practically on actual
$t$ dependence of the phase $\zeta^N(s,t)$
used in the analysis. 
  
The values of the root-mean-square of inelastic impact parameter 
calculated with the help of Eq. (\ref{mv13}) are 
significantly smaller 
in the peripheral case of elastic hadron scattering 
than in the central case. In the case of central behavior
of elastic hadron scattering the root-mean-square 
($\sim$ 0.68 fm) of the
elastic impact parameter is even 
lesser then the root-mean-square
of the inelastic one ($\sim$ 1.1 fm); the elastic 
hadron scattering being more central
(see also Ref. \cite{miet})
than inelastic scattering processes.
The situation is reversed in the case 
of peripheral elastic hadron profiles.
The more pronounced centrality of 
the deep inelastic scattering processes, the higher peripherality
of elastic hadron scattering.
Figs. 4, 5 exhibit the graphs 
of the root-mean-squares for
the total, elastic and inelastic impact parameters calculated
with the help of Eqs. (\ref{mv8}), (\ref{mv3}) and 
(\ref{mv13}) for corresponding $\rho$ values.

Further analysis of the $pp$ elastic hadron scattering
at this energy shows that also 
in the case of higher peripherality, characterized
by the higher values of the elastic root-mean-squares,
e.g., $\sqrt {<b^2>_{el}} \sim 1.95 \div 1.98$ fm,
the region of admissible $\rho$ values is not
changed substantially (and of $\sigma_{tot}$ and $B$, too). 

The typical $t$ dependence of elastic hadron phase
$\zeta^{N} (s,t)$ is shown in Fig. 6.
It may be seen that there is a fundamental 
difference in the $t$ dependence for 
peripheral (full line) and the central (dashed line)
behaviors. Corresponding impact parameter profiles
are shown in Fig. 7.

\subsection{$\bar{p} p$ elastic scattering at energy of 541 GeV}

The data for the analysis of $\bar{p} p$ elastic
hadron scattering at energy of 541 (546) GeV
were taken from Refs. \cite{aug1,bozz} 
(data at energy of 630 GeV \cite{bern} being also included).
The data in individual measured kinematical regions
were normalized in a similar way as in the previous
$pp$  case. Moreover, the normalization condition, 
used in UA4/2 experiment \cite{aug1}, i.e.,
$\sigma_{tot} (1 + \rho^2) = 63.3 \pm 1.5 $ mb, was 
also taken into account.

The numerical values of basic physical characteristics
of the elastic hadron scattering amplitude, i.e., 
the total cross section, the diffractive slope and the  
quantity $\rho$ are shown in lower part of Table I. 
The results of standard analysis  
obtained with the simplified form of WY total elastic 
scattering amplitude (see Ref. \cite{aug1})
are also included.

Fig. 8 shows the $\chi ^2$ distributions
for different $\rho$ values in both the peripheral 
and central cases where the data from the total
measured interval of momentum transfers were taken
into account. Fig. 9 shows
the values of the total cross section
(obtained with the help of Eq. (\ref{st1}))
corresponding to different $\rho$ values.
The condition $\sigma_{tot} (1 + \rho^2) = 63.3 \pm 1.5$  mb
used in the normalization of experimental data at small $|t|$
\cite{aug1} limits significantly the region of admissible total
cross section values; it admits for the $\rho$ 
to be in the interval (0.11, 0.18) 
in the peripheral case; and within the 
interval (0.08, 0.14) in the central case.

However, the interval of admissible $\rho$ values
in the peripheral case would be broader,
i.e., (0.11, 0.23),
if instead of the previous normalization
condition the value of $\sigma_{tot} = 63.0 \pm 2.1$ mb
were used; the value of total cross section 
being estimated with the help of another
(luminosity dependent) method  \cite{aug2}.
In the central case the value of $\rho$ would 
lie within the interval (0.08, 0.16).

Fig. 10 shows the values of the diffractive slope B
(determined with the help of Eq. (\ref{sl1}) 
in forward direction) corresponding to different
$\rho$ values. The used normalization condition 
limits its values to the interval (16.20, 16.55) GeV$^2$
in the peripheral case.
In the central case its value would 
lie within the interval (15.80, 16.00).
And finally the dependence of
the root-mean-squares of the total, elastic 
and inelastic impact parameters 
for different $\rho$ values are shown in Figs. 11 and 12. 
The root-mean-square values of the total impact
parameter determined with the help of Eq. (\ref{mv8}) 
equal approximately 1.1 fm (they are a little bit higher than
in the $pp$ case).

As it was mentioned, the value of the root-mean-square
of elastic impact parameter calculated with the help of 
Eq. (\ref{mv3}) is composed of two terms.
The first term representing the contribution
of the modulus of the elastic hadron amplitude
gives the value of 0.76 fm;
a little bit higher than in the previous $pp$ case. 
The second term, containing the contribution of the 
phase of elastic hadron amplitude, reaches
in a peripheral case the value about 2.1 fm
and is again practically zero for 
the central behavior.  
The values of the root-mean-square of inelastic impact parameter 
calculated with the help of Eq. (\ref{mv13}) are 
in the peripheral case of elastic hadron scattering
significantly smaller: $\sim 0.4 \div 0.5$ fm ;
see Table II. The $t$ dependence of the phase in 
peripheral and central cases is represented in Fig. 13;
and corresponding elastic profiles in the impact
parameter space in Fig. 14.
 
\section{Actual inelastic and total profiles}

The function $c(s,b)$ enabling to
derive the densities of total and inelastic scatterings
in the impact parameter space has been introduced
in Sect. 2. It can be hardly derived analytically; however, 
its numerical values can be derived with the help of
numerical procedure. Its $b$ dependence 
found in such a way for the peripheral
case of $\bar{p} p$ elastic scattering at energy of 541 GeV
is presented in Fig. 15. The corresponding total 
and inelastic profiles are shown, too.
They are both of central character, while 
the original elastic profile remains 
unchanged and is peripheral.
Also values of the total and
inelastic root-mean-squares derived
from the $t$ dependent elastic hadron 
scattering amplitude are reproduced 
by these newly established total and inelastic profiles. 

Fig. 16 shows the shape of correction
function $K(s,b)$ calculated from the $t$ dependent elastic
hadron scattering amplitude $F^N(s,t)$ with the help
of formula (\ref{cf1}). It corresponds to the peripheral 
picture of elastic $\bar{p} p$ hadron scattering. 
The absolute values of $K(s,b)$ at the
given value of $b$ are about 19 orders of magnitude
lesser compared to the remaining quantities
in unitarity equation (\ref{mv14}) at the same $b$
and can be, therefore, practically
neglected.

\section{Transparent or hard nucleon?}

The $pp$ elastic scattering at energy of 53 GeV
was analyzed by Miettinen \cite{miet}
provided that the imaginary part of elastic hadron
scattering amplitude is dominant in a broad interval
of $t$. Having used the FB transform and unitarity
condition he established elastic, inelastic and
total profiles in the impact parameter space; 
all these profiles being central.
Comparing the inelastic profile to that 
corresponding to the black disc model 
he concluded that for $b = 0$ there is cca 6 $\%$ of events
in which no inelastic (absorption) process occurs;
the value being regarded very high from the point of
realistic conditions.
Consequently, the nucleon was claimed by Miettinen to
be a transparent object enabling to exhibit
elastic scattering even in central collisions.

The approach of Miettinen was repeated by
Henzi and Valin \cite{henz} 
and applied to elastic hadron $\bar{p} p$ 
scattering at energy of 546 GeV.
They obtained that nucleons at this energy
should be more opaque with elastic impact
parameter profile being more edgier and having a greater
range than in the case of ISR energies.

It is evident that the idea of transparent
nucleon (having influenced significantly
all discussions concerning quark structure of hadrons)
has followed from the assumption of dominant
imaginary part of elastic amplitude, representing
an {\it{a priori}} strongly limited condition. 
There is not any need of transparency if 
elastic collisions are described with the help of
a more exact formula and are allowed
to be peripheral. In such a case the
root-means-square of elastic impact parameter
(for $pp$ elastic scattering at energy of 53 GeV,
e.g., 1.78$\div$ 1.80 fm, and for $\bar{p} p$
scattering at energy of 541 GeV, e.g.,
2.24 $\div$ 2.35 fm) are significantly higher
than these for inelastic processes (i.e., 
0.68 or 0.76 fm in these two cases - see Table II).

Peripheral behavior is preferred by the total
$\chi^2$ values being significantly
lower than for any alternatives requiring
the central behavior. There is not,
therefore, any reason to regard nucleons 
in elastic high-energy collisions
as transparent objects.

However, the nucleons cannot be denoted as 
standard (classical) matter objects having a fixed 
dimension in the transverse direction
at any time. On the basis of our analysis
they can be regarded as objects with
a rather hard core of diameter cca  0.4 $\div$ 0.8 fm
and with a practically transparent outer shell of
diameter 1.8 $\div$ 2.4 fm; the given core being
responsible for inelastic processes and
the outer shell for elastic scattering (or perhaps
for some diffractive production processes, too).
The value of core diameter is a little bit
smaller than the proton charge diameter determined
with the help of different charge distribution
methods inside proton (see, e.g., Ref. \cite{kars}).
 
Such a collision structure may result 
from two different reasons. First, all 
experimental data must be regarded as an
average over divers spin orientations. 
And a transverse momentum may correspond to
different nucleon polarizations in individual events.
The other reason might be related to the 
possibility that the dimensions of nucleons
consisting of internal moving partons need 
not to be quite constant in all time. Experiments with
polarized nucleons might contribute
to the solution of this problem.

\section{Conclusion}

The new formulas describing elastic hadron 
scattering and presented in this paper
have enabled to establish the impact
parameter profiles for elastic and inelastic
processes. The values of corresponding 
root-mean-squares may be taken as
characteristics of the ranges of hadron forces
responsible for elastic and inelastic collision processes.

The corresponding numerical values were derived from 
experimental data of elastic $pp$ scattering
at energy of 53 GeV and $\bar{p} p$
scattering at energy of 541 GeV. They 
depend to some extent (if not substantially) on the
peripherality degree imposed on the
elastic profiles in establishing 
the $t$ dependence of elastic hadron scattering amplitude
(especially, of its phase) from the experimental data.
Higher peripherality degrees are preferred by total
$\chi^2$ values, which indicates that there is not any reason
to regard protons as transparent objects in elastic high-energy
collisions.

The presented analysis also leads to the conclusion
that it is not sufficient to characterize high-energy
elastic collisions only by usual three quantities:
total cross section $\sigma_{tot}$,
diffractive slope $B$ and the ratio $\rho$
of real to imaginary parts of elastic hadron
amplitude in the forward direction.
All these quantities depend on
peripherality degree imposed,
significant dependence being exhibited especially by the 
quantity $\rho$. The presented results
have opened a series of new questions 
concerning the actual structure  of hadrons 
playing a role in collisions of different types.

\makeappendix{: Exact impact parameter  representation of 
scattering amplitude}

The approach used in Refs. \cite{adac} and \cite{isla}
starts from the possibility of defining the amplitude 
$A(s,t)$ being identical with the 
elastic hadron amplitude $F^{N}(s,t)$ in the physical $t$ region,
while in the unphysical region of $t$ the amplitude $A(s,t)$
equals unknown complex function $\lambda (s,t)$ which is assumed 
to be of a bounded variation for all $t$ values from
$-\infty < t < t_{min}$ and which further fulfilles 
condition that the integral
\begin{equation}
\int \limits_{- \infty}^{t_{min}} \! \lambda (s,t) (-t)^{-1/4} dt
\label{lk1}
\end{equation}
is absolutely convergent. Both the conditions guarantee (according to 
Hankel's theorem \cite{wats}) that $A(s,t)$ has a Fourier-Bessel
representation for $ -\infty < t < 0$.

The elastic scattering amplitude $h_{el}(s,b)$
can be then expressed as a sum of two terms
\begin{equation}
h_{el}(s,b) = h_1 (s,b) + h_2 (s,b),
\label{el1}
\end{equation}
where
\begin{equation}
h_1 (s,b) = {1 \over {4 p \sqrt{s}}} 
\int \limits_{t_{min}}^{0} \! dt \; F^{N}(s,t)\; J_{0}(b \sqrt{-t}),
\label{el2}
\end{equation}
and
\begin{equation}
h_2 (s,b) = {1 \over {4 p \sqrt{s}}} 
\int \limits_{-\infty}^{t_{min}} \! dt \; \lambda (s,t) \; J_{0}(b \sqrt{-t}).
\end{equation}

One can divide the inelastic overlap
function into two parts in the same way \cite{isla}
\begin{equation}
g_{inel}(s,b) = g_1 (s,b) + g_2 (s,b),
\end{equation}
where
\begin{equation}
g_1 (s,b) = {1 \over {4 p \sqrt{s}}} 
\int \limits_{t_{min}}^{0} \! dt \; G_{inel} (s,t)\; J_{0}(b \sqrt{-t}),
\label{gn1}
\end{equation}
and
\begin{equation}
g_2 (s,b) = {1 \over {4 p \sqrt{s}}} 
\int \limits_{-\infty}^{t_{min}} \! dt \; \mu (s,t) \; J_{0}(b \sqrt{-t}).
\label{gn2}
\end{equation}
The real function $\mu (s,t)$ introduced in 
Eq. (\ref{gn2}) is restricted by the same
two conditions as function $\lambda (s,t)$.

It also holds \cite{adac}
\begin{equation} 
\int \limits_{0}^{\infty} b db \; \Im h_{2}(s,b) = 
\int \limits_{0}^{\infty} b db \; g_{2}(s,b) =0.
\label{cs3}
\end{equation}

Consequently, the following unitarity equation
in the impact parameter space  can be written
which binds together only the
physically relevant amplitudes $h_{1}(s,b)$ 
and  $g_{1}(s,b)$ \cite{adac,isla}:
\begin{equation}
\Im h_{1}(s,b) = |h_{1}(s,b)|^2 + g_{1}(s,b)  + K(s,b).
\label{mv14}
\end{equation}

The function $K(s,b)$ can be determined with the help 
of elastic hadron amplitude $F^{N}(s,t)$  by
\begin{eqnarray}
K(s,b) &=& { 1 \over {16 \pi^2 s}}\int \limits_{t_{min}}^{0} \! dt_1 \!
 \int \limits_{t_{min}}^{0}\! dt_2 \;
F^{N*}(s,t_2) \; F^{N}(s, t_1)
  \bigg [ J_{0} \bigg ({b \over {2p}} \sqrt {-t_1 (4 p^2 + t_2)} \bigg )
\nonumber \\  
 &&       J_{0} \bigg ({b \over {2p}} \sqrt {-t_2 (4 p^2 + t_1)} \bigg )       
         - J_{0} (b \sqrt{-t_1}) \; J_{0} (b \sqrt{-t_2}) \bigg ].
\label{cf1}
\end{eqnarray}
The function $K(s,b)$ vanishes at 
$b = 0$ and $b \rightarrow \infty$.

Adachi and Kotani \cite{adac}
showed then that contrary to general
belief the functions $\Im h_{1}(s,b)$ and 
$g_{1}(s,b)$ cannot be non-negative at all
finite positive $b$ at finite $s$. It must hold only 
$\Im h_{1}(s,0) \geq 0$ and both the functions
$\Im h_{1}(s,b)$ and $g_{1}(s,b)$ 
must vanish when $b$ tends to $\infty$.  
At higher finite values of $b$
these functions should oscillate.
The oscillations vanish at infinite energies only.

Multiplying the both sides of Eq. (\ref{mv14}) by 
$8 \pi b$ and integrating 
over the all possible impact parameter values $b$
one obtaines in our normalization 
\begin{equation}
\sigma_{tot}(s) = \sigma_{el}(s) + \sigma_{inel}(s) +
{8 \pi}\int \limits_{0}^{\infty} b db \; K(s,b),
\end{equation}
which shows that the function $K(s,b)$ 
fulfills the condition:
\begin{equation}
\int \limits_{0}^{\infty} b db \; K(s,b) = 0.
\end{equation}
Here $\sigma_{tot}(s)$, $\sigma_{el}(s)$ and $\sigma_{inel}(s)$
are the total, elastic and inelastic cross sections
defined as integrals
\begin{equation} 
\sigma (s) =
{8 \pi } \int \limits_{0}^{\infty} b db \; Z(s,b),
\label{cs1}
\end{equation}
where $Z(s, b)$ stands for one of the amplitudes
$\Im h_{el}(s,b)$, $|h_{el}(s,b)|^2$ or
$g_{inel}(s,b)$.

All the integrals $\sigma (s)$ in Eq. (\ref{cs1})
have the physical meaning as introduced earlier
but only $|h_{1}(s,b)|^2$ is non-negative at any $b$.
Thus, only $|h_{1}(s,b)|^2$ can be interpreted as 
the density distribution (i.e., of elastic hadron scattering) 
in the impact parameter space. 
The other functions $\Im h_{1}(s,b)$ and $g_{1}(s,b)$ 
can turn to oscillate at some higher values of $b$;
therefore, they cannot be interpreted directly
as corresponding density distributions.

However, they can be modified to fulfil such a goal.
Adding a suitable real function $c(s,b)$,
fulfilling the  condition
\begin{equation}
\int \limits_{0}^{\infty} b db \; c(s,b) = 0,
\label{ca1}
\end{equation}
to the both sides of unitarity equation (\ref{mv14}),
both the functions $h_{tot}(s,b)=\Im h_1(s,b) + c(s,b)$ 
and $g_{inel}(s,b)=g_1(s,b) + K(s,b) + c(s,b)$ 
can be brought to be non-negative
for all the values of $b$. 
Such a function $c(s,b)$ exists owing to
the properties of function
$g_2 (s,b)$, or equivalently owing to
properties of the function 
$\mu (s,b)$ defined in the unphysical
region of $t$. 
Then the functions $h_{tot}(s,b)$ 
and $g_{inel}(s,b)$ can be regarded as new density
distributions of the total and inelastic 
scatterings in the impact parameter space and
one has instead of unitarity equation (\ref{mv14})
the modified unitarity condition
\begin{equation}
h_{tot}(s,b) = |h_{1}(s,b)|^2 + g_{inel}(s,b).
\label{mv15}
\end{equation}
Thus, we have in principle a fully consistent description of
elastic scattering in the impact parameter space.

\newpage

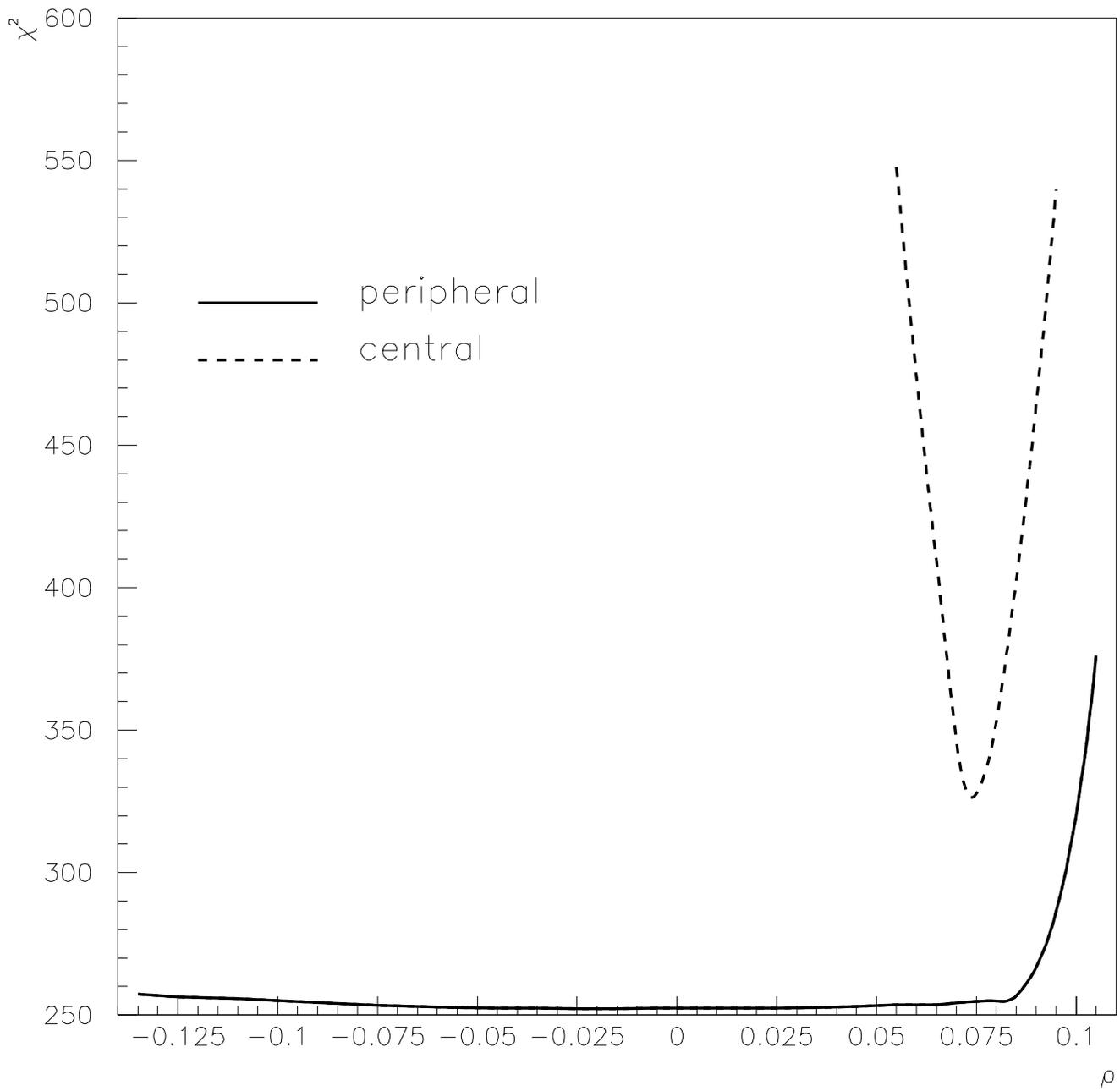
\begin{figure}
\caption{Dependence of the total $\chi^2$ values upon 
         the chosen value of $\rho$ for
         $pp$ elastic scattering at energy of 53 GeV}
\end{figure}

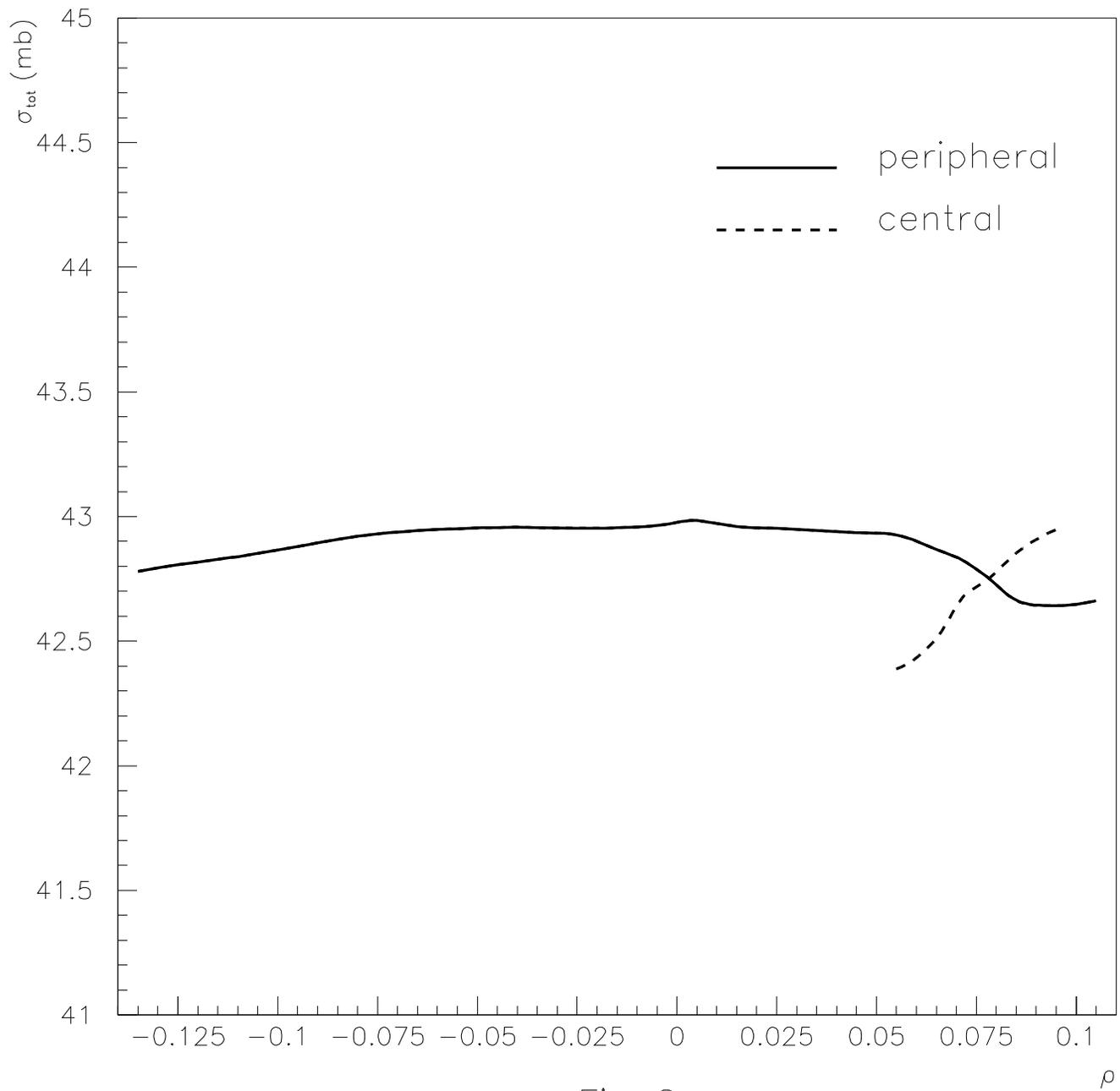
\begin{figure}	 
\caption{Dependence of the total cross section in
         $pp$ elastic scattering on the chosen value of $\rho$
         at energy of 53 GeV}
\end{figure}

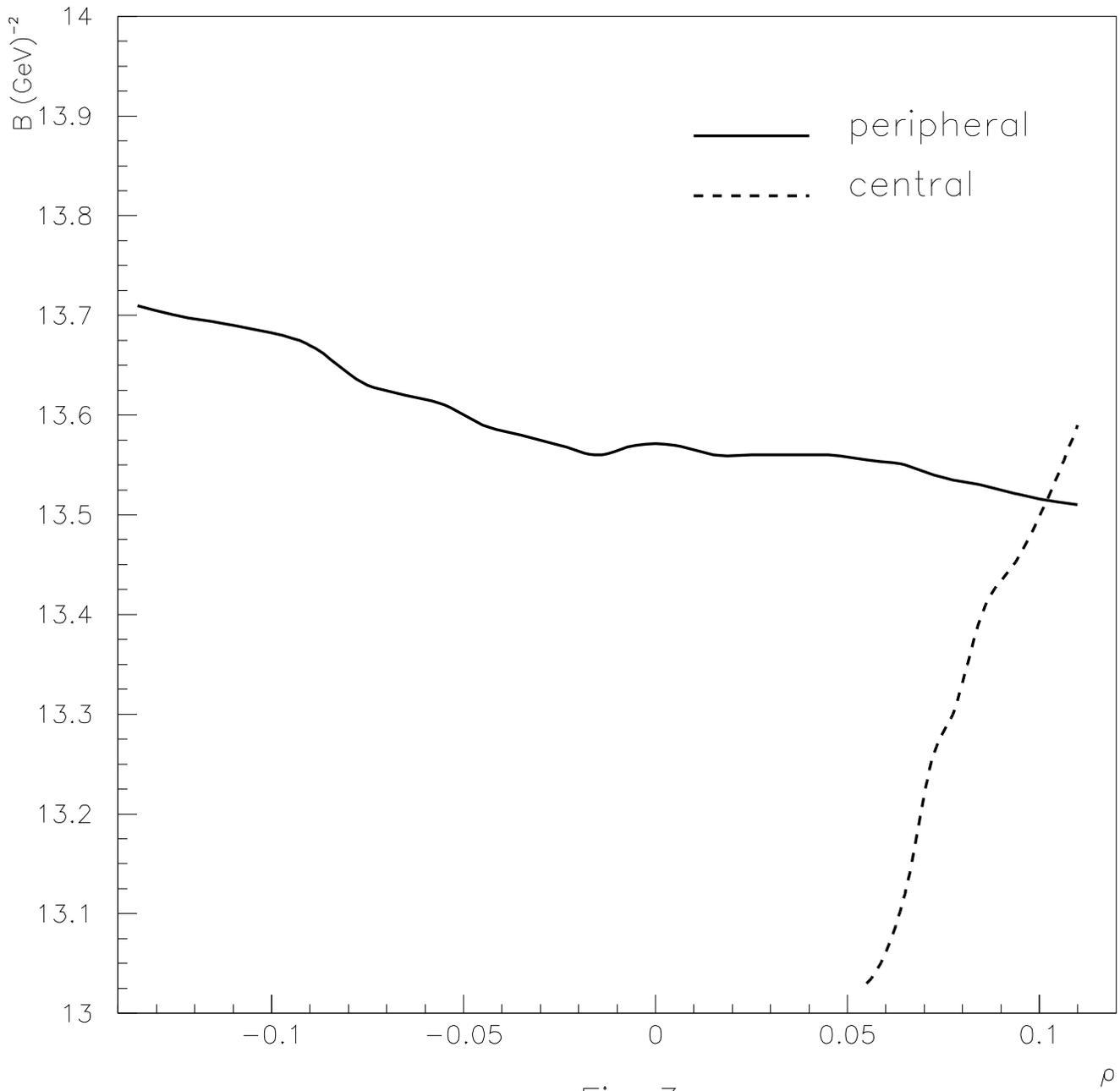
\begin{figure} 
\caption{Dependence of the diffractive slope
         $B$ values upon $\rho$ for
         $pp$ elastic scattering at energy of 53 GeV}
\end{figure}

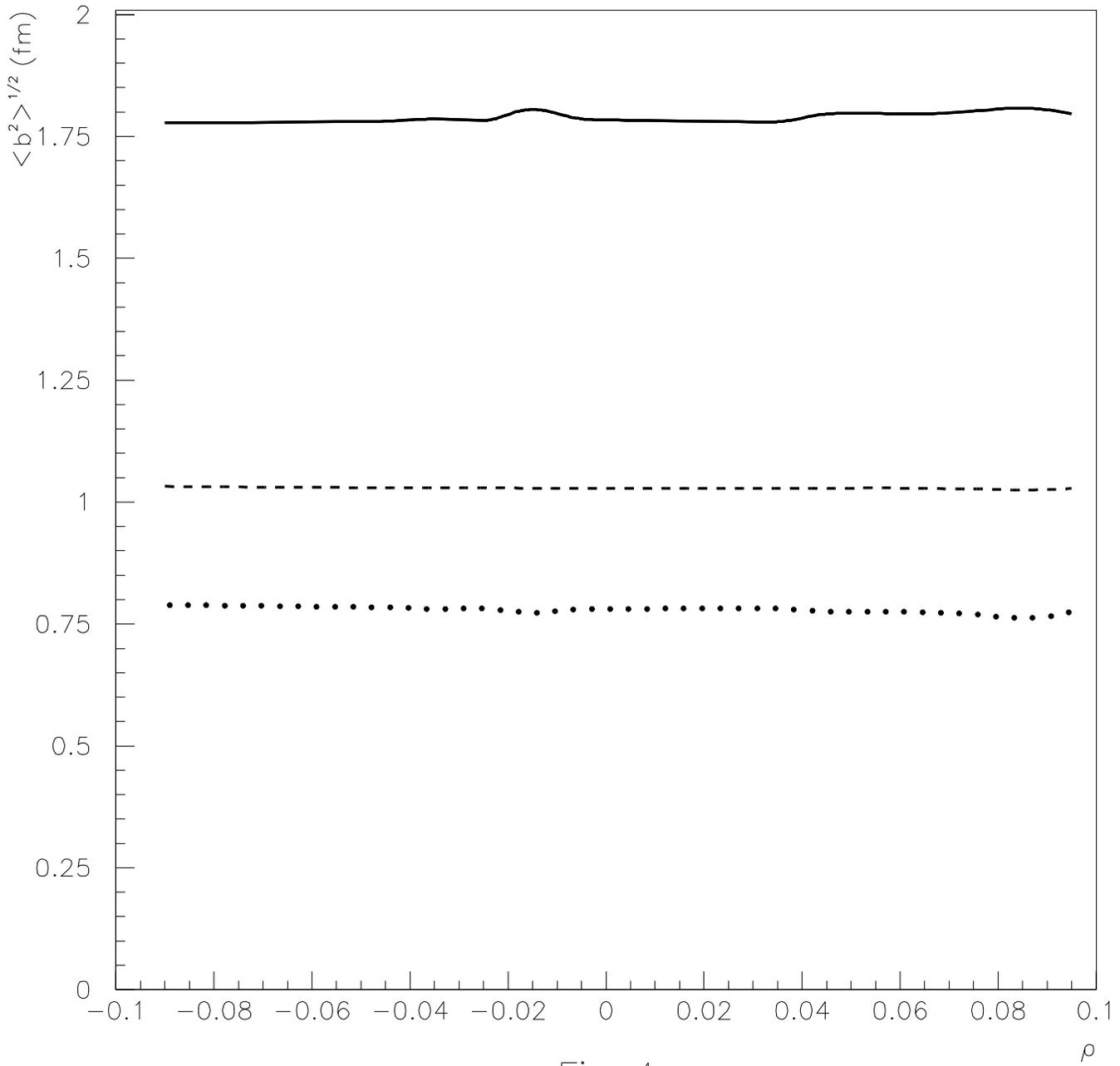
\begin{figure}
\caption{Dependence of the root-mean-squares
         for the total (dashed line), elastic (full line)
         and inelastic (dotted line) impact parameters
         upon the chosen value of $\rho$;
         $pp$ elastic scattering at energy of 53 GeV -  
         peripheral case}
\end{figure}

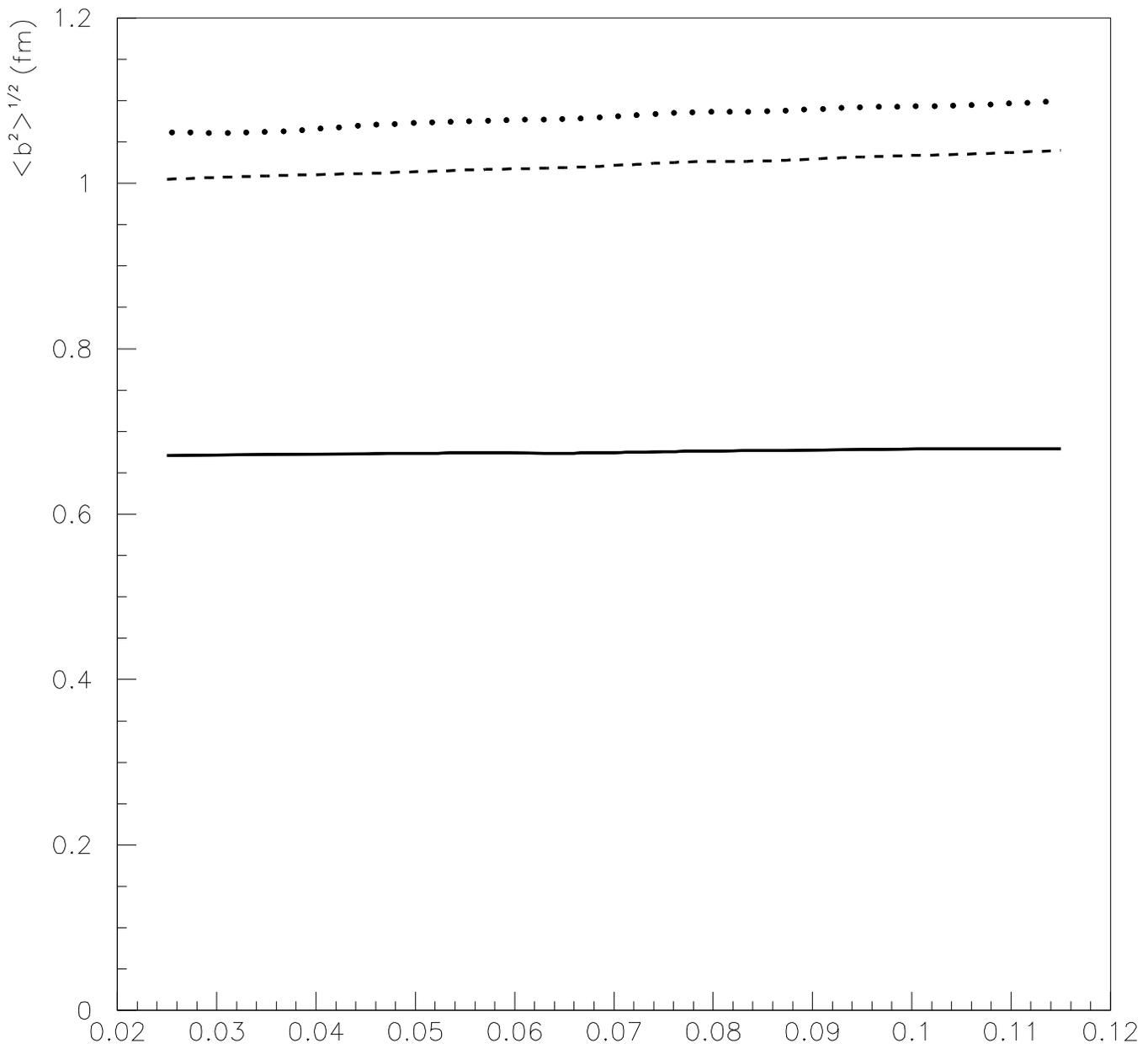
\begin{figure}       
\caption{Dependence of the root-mean-squares
         for the total (dashed line), elastic (full line)
         and inelastic (dotted line) impact parameters
         pon the chosen value of $\rho$;
         $pp$ elastic scattering at energy of 53 GeV -
	 central case}
\end{figure}

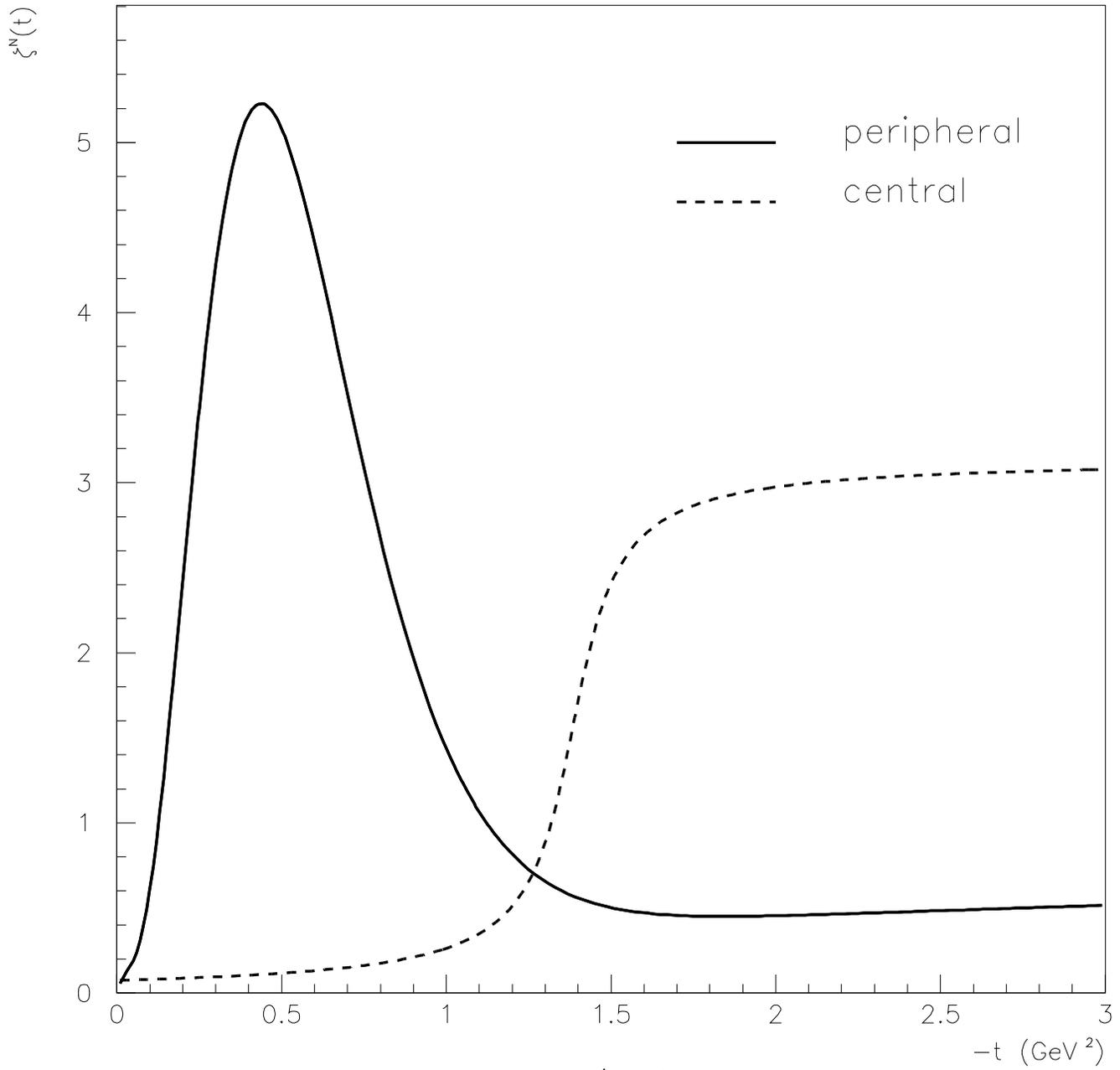
\begin{figure}	 
\caption{The $t$ dependence of the phase corresponding to
         the peripheral and the central cases
         of $pp$ elastic hadron scattering at energy of 53 GeV}
\end{figure}

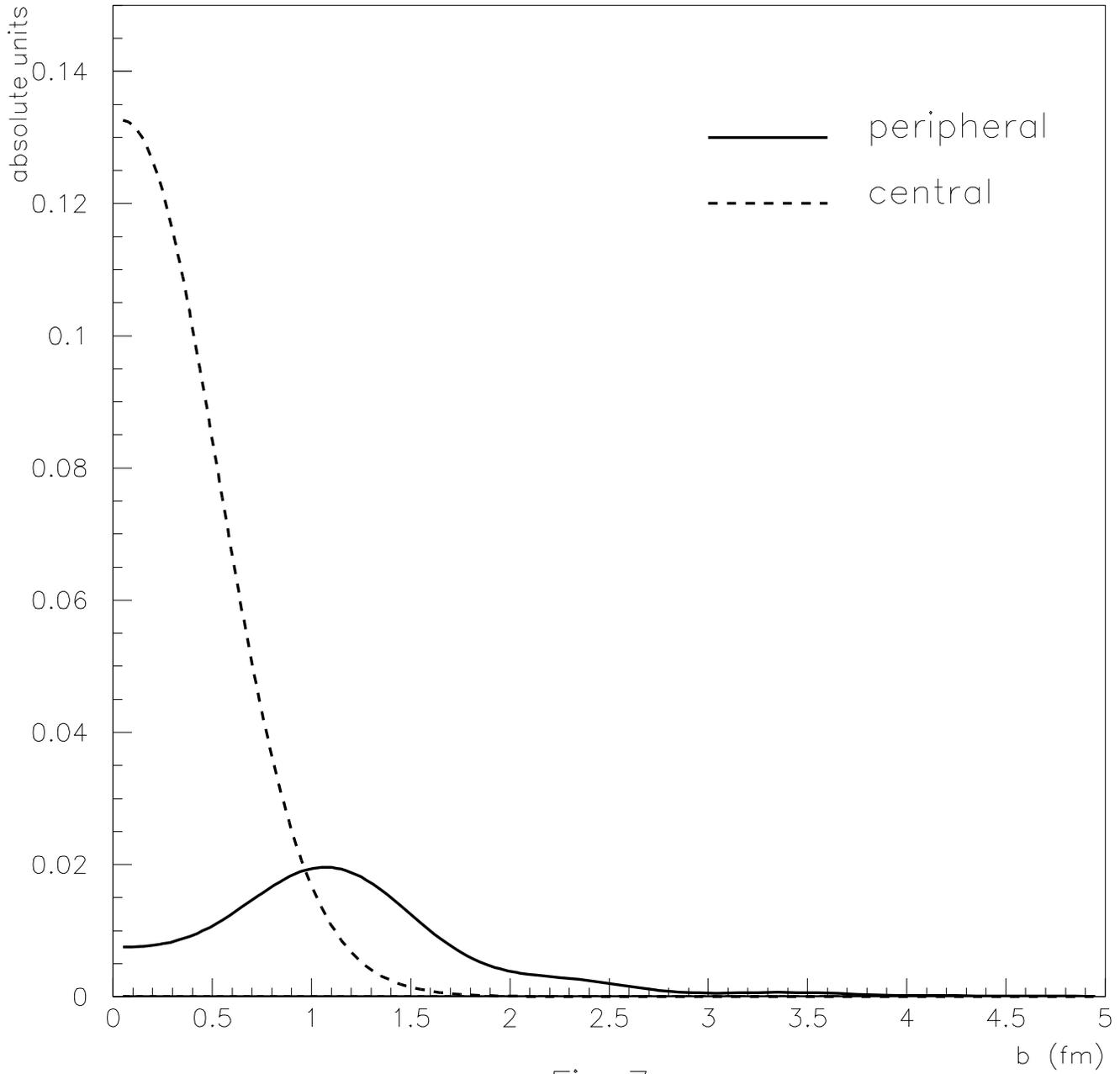
\begin{figure}	      
\caption{Elastic impact parameter profiles in the 
         peripheral and central cases of
         pp elastic scattering at energy of 53 GeV}
\end{figure}

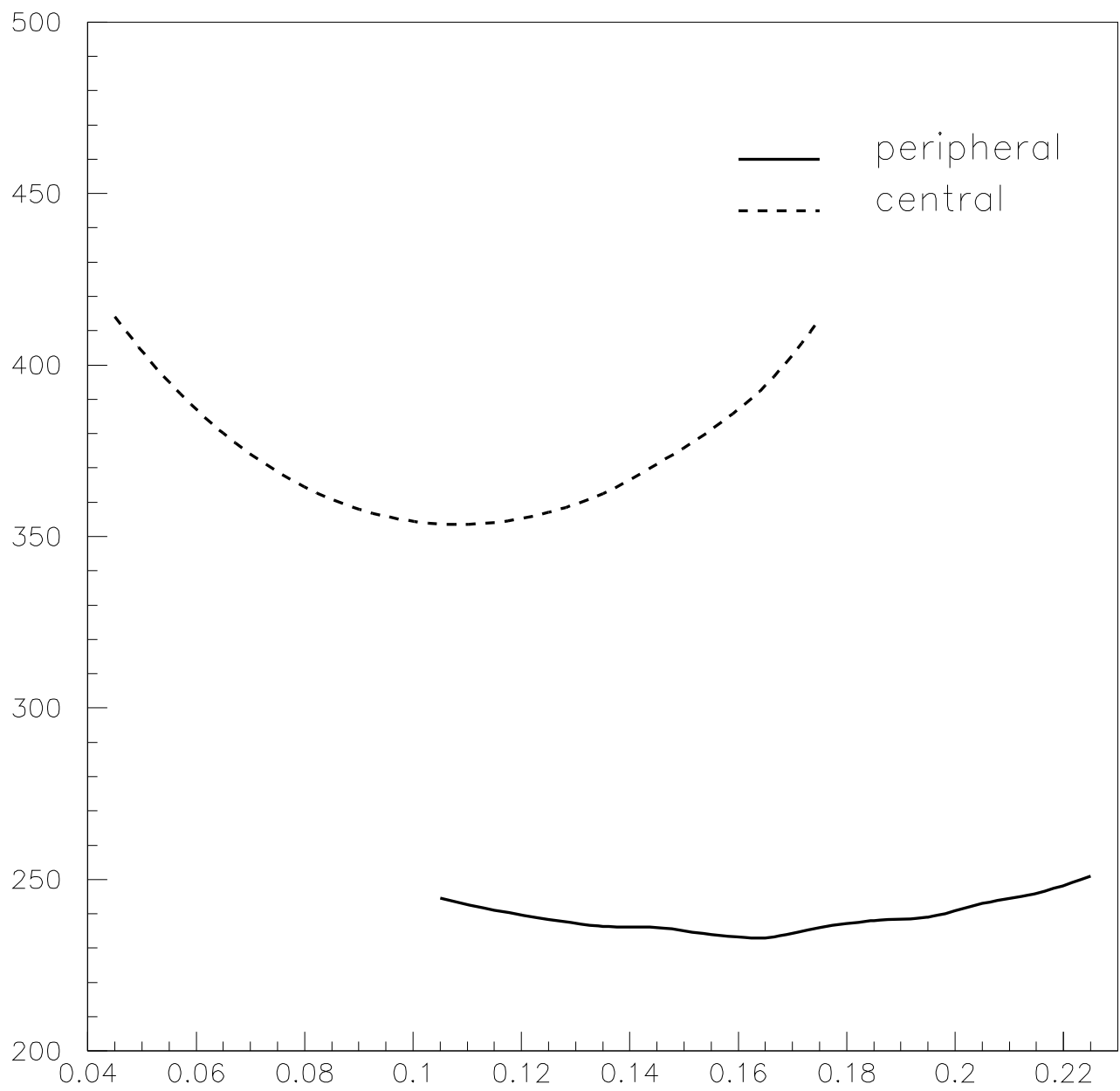
\begin{figure}	 
\caption{Dependence of the total $\chi^2$ values upon $\rho$ for
         $\bar{p} p$ elastic scattering at energy of 541 GeV}
\end{figure}

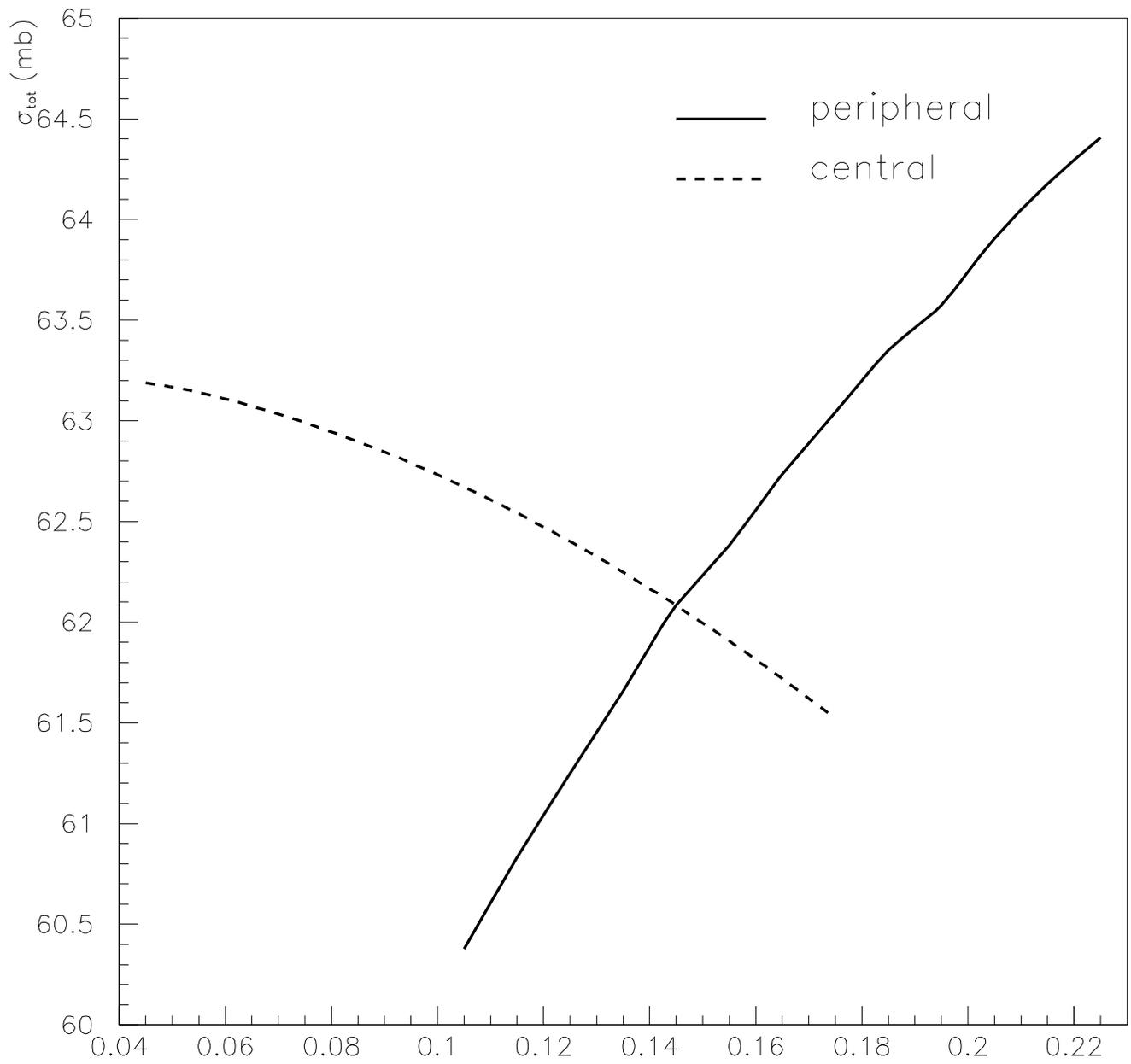
\begin{figure}	 
\caption{Dependence of the total cross section
         $\sigma_{tot}$ values upon $\rho$ for
         $\bar{p} p$ elastic scattering at energy of 541 GeV}
\end{figure}

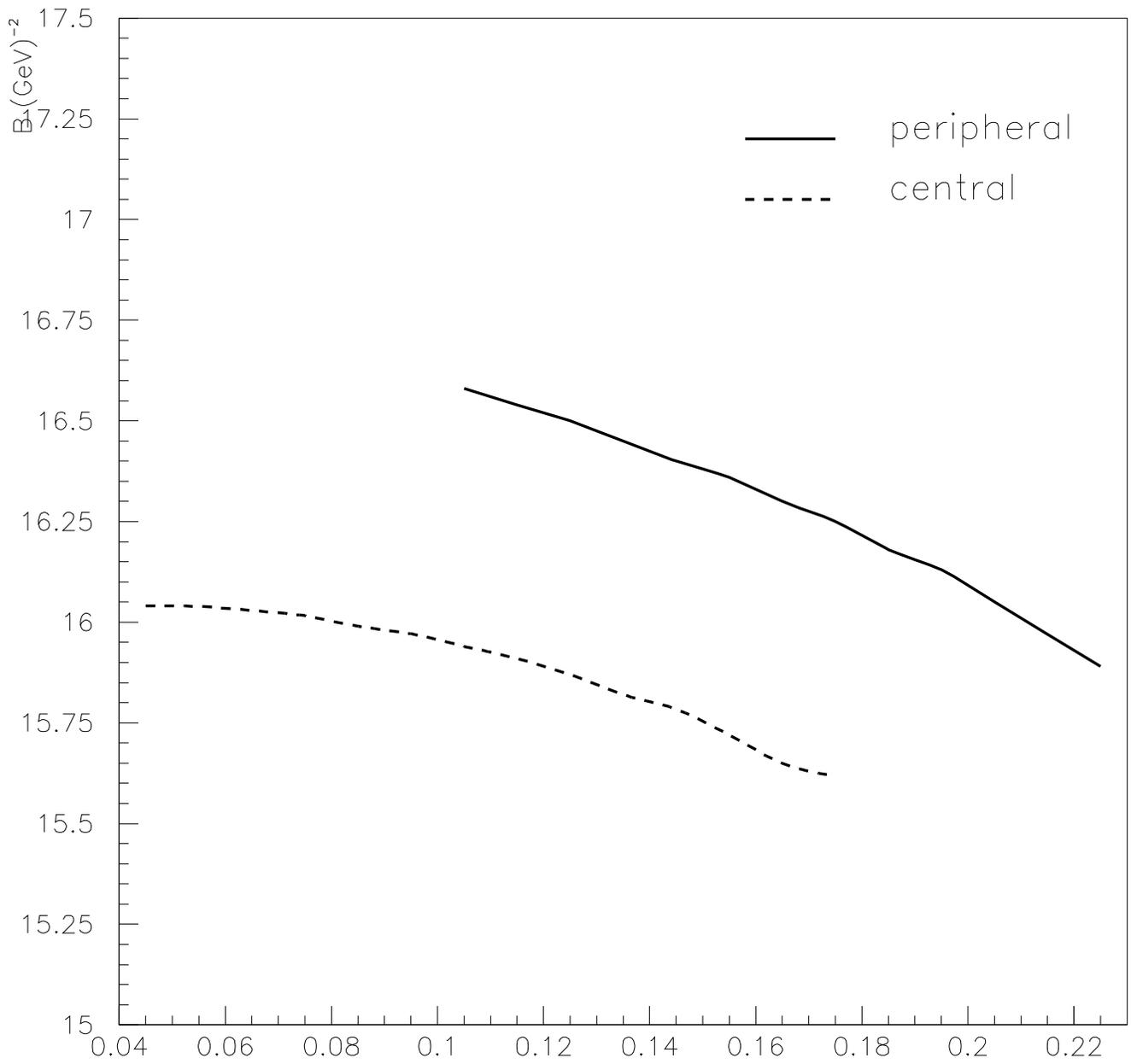
\begin{figure}	 
\caption{Dependence of the diffractive slope
         $B$ values upon $\rho$ for
         $\bar{p} p$ elastic scattering at energy of 541 GeV}
\end{figure}

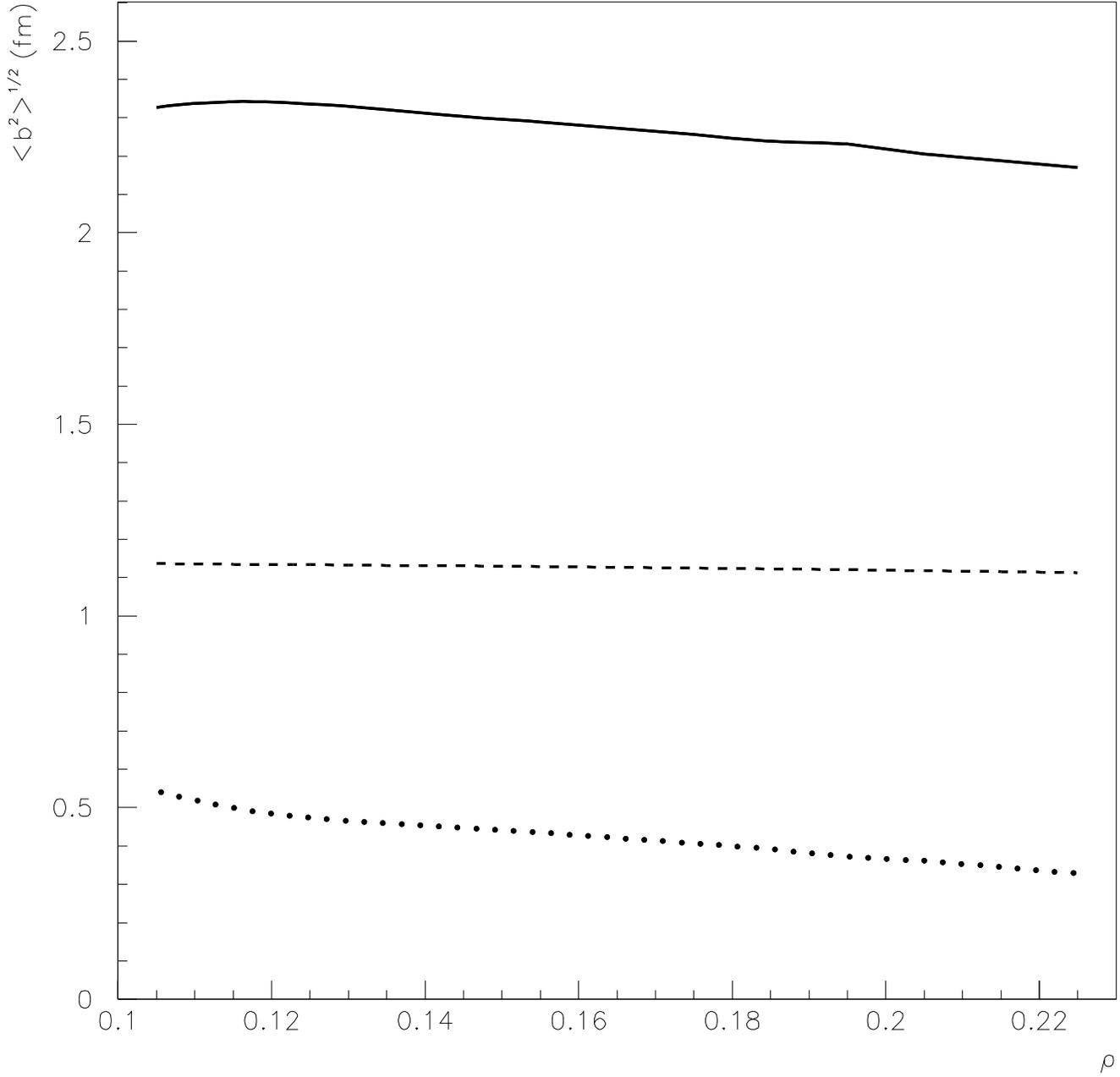
\begin{figure}	 
\caption{Dependence of the root-mean-squares
         for the total (dashed line), elastic (full line)
         and inelastic (dotted line) impact parameters
         upon $\rho$ for $\bar{p} p$ elastic scattering at energy of 541 GeV -
         peripheral case}
\end{figure}

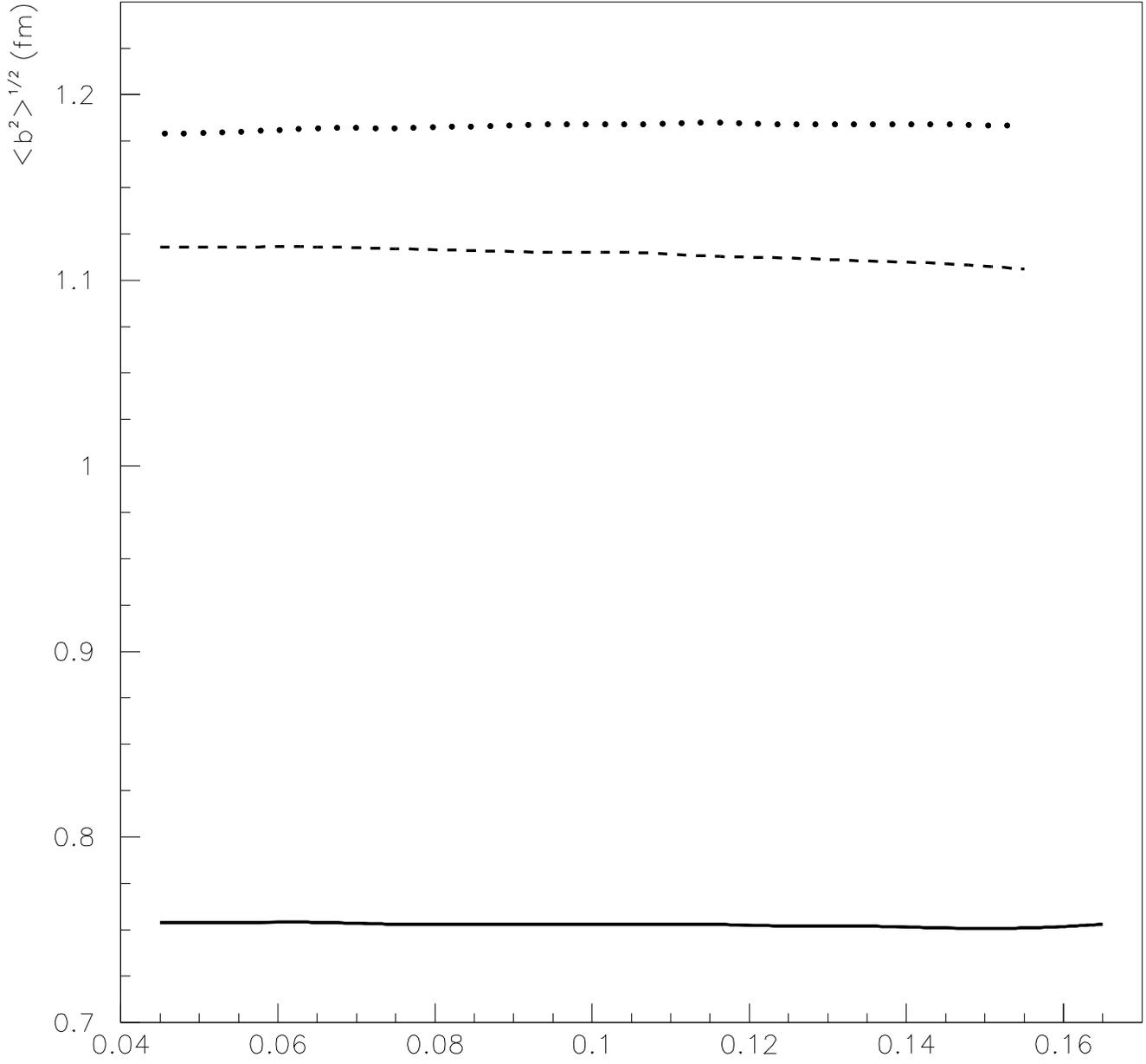
\begin{figure}	 	   
\caption{Dependence of the root-mean-squares
         for the total (dashed line), elastic (full line)
         and inelastic (dotted line) impact parameters
         upon $\rho$ for $\bar{p} p$ elastic scattering at energy of 541 GeV -
         central case}
\end{figure}

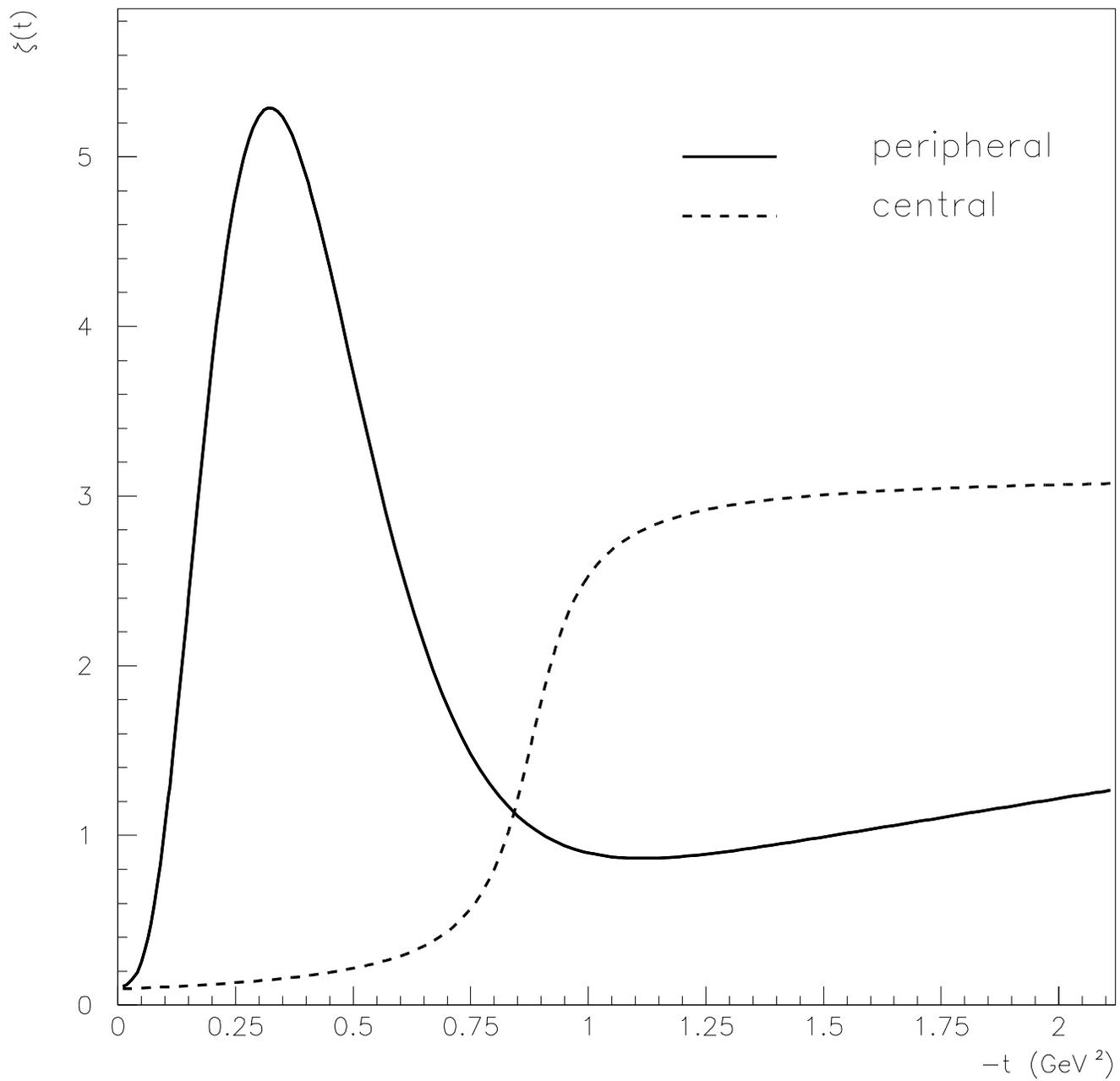
\begin{figure}	  
\caption{$t$ dependence of the phase in the peripheral
         and central cases; $\bar{p} p$
         elastic scattering at energy of 541 GeV}
\end{figure}
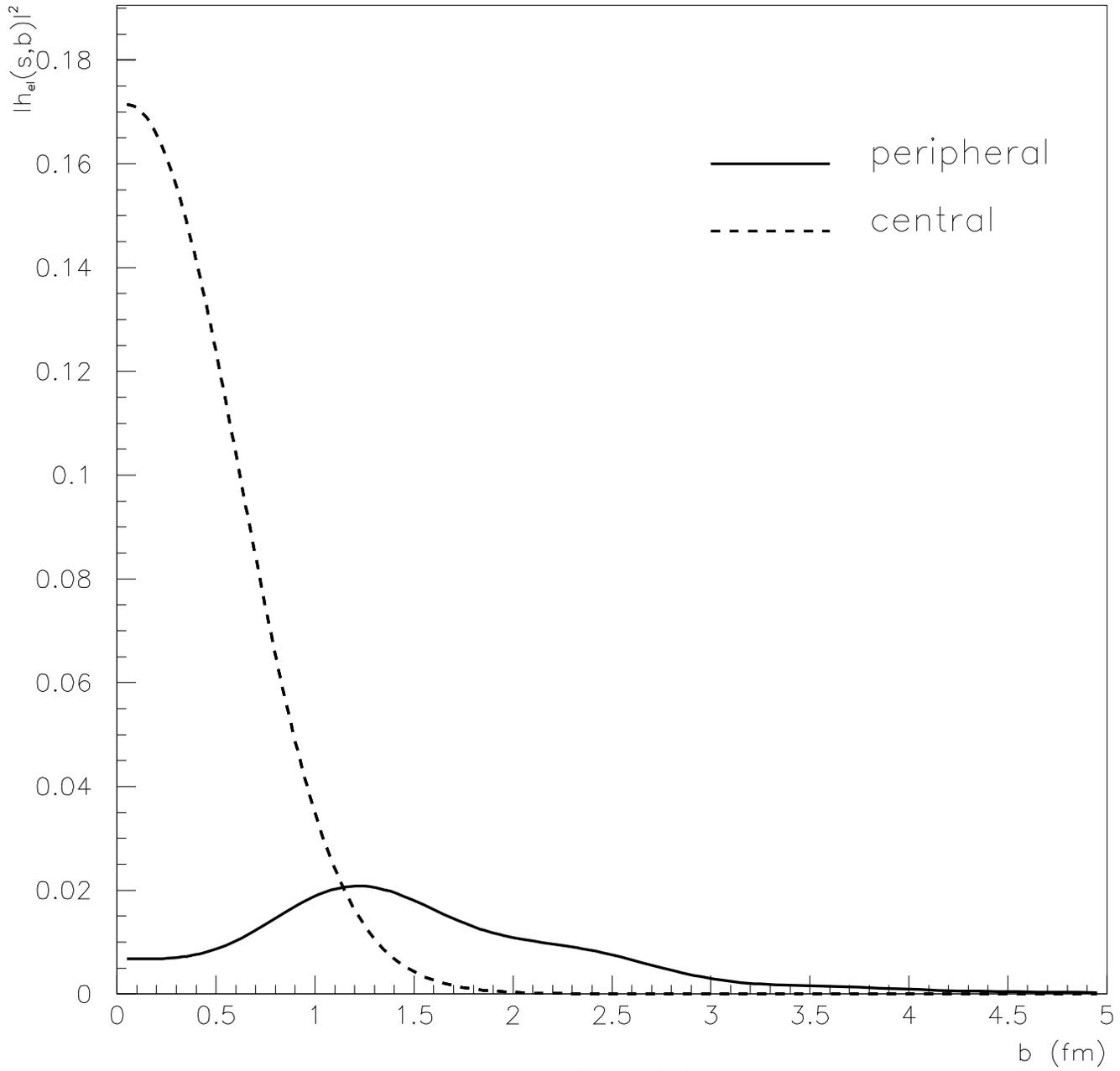
\begin{figure}	 
\caption{Impact parameter distribution of elastic $\bar{p} p$ 
         elastic scattering at energy of 541 GeV}
\end{figure} 	 	   
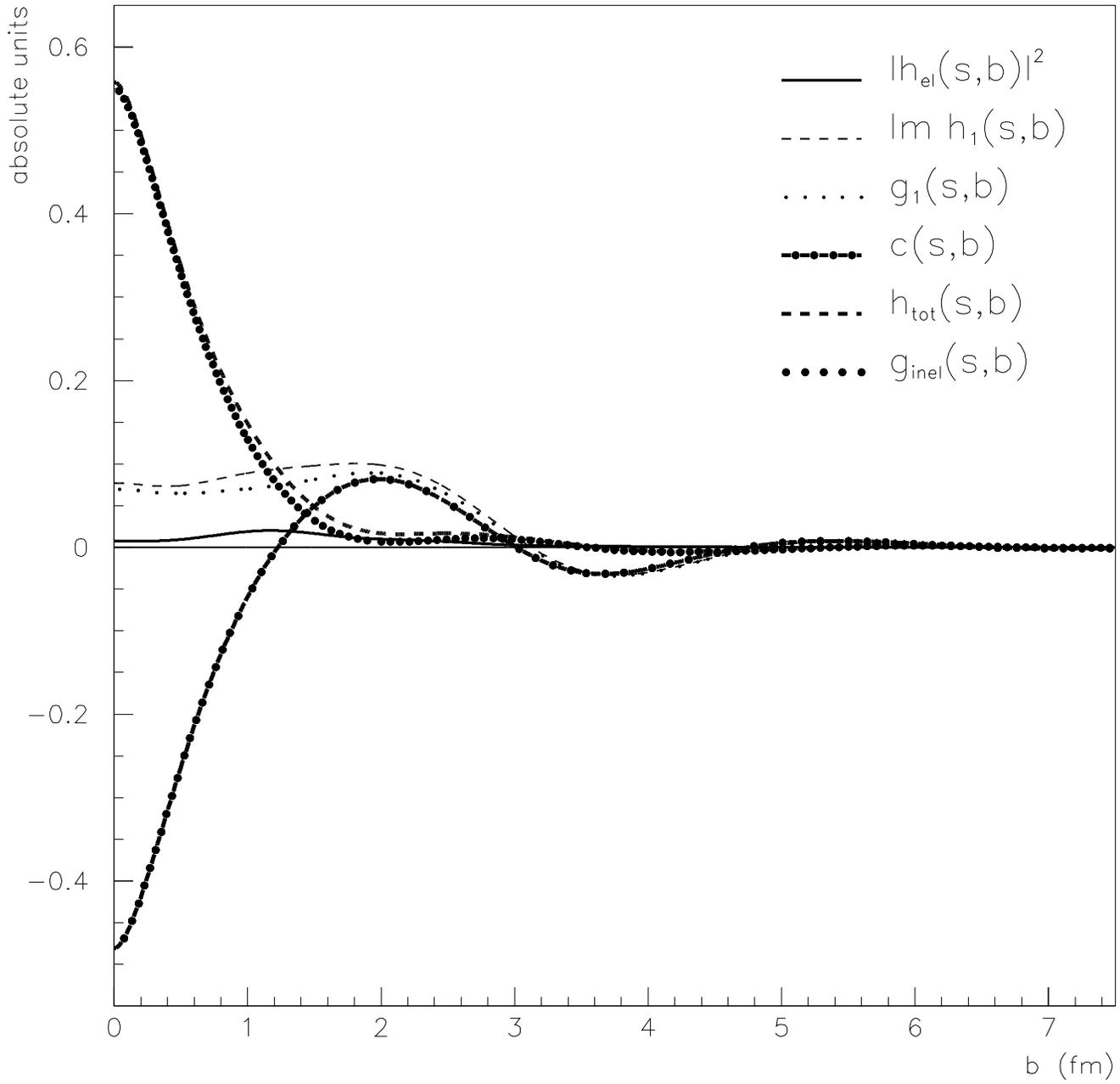
\begin{figure}
\caption{New profiles constructed with the help of
         function $c(s,b)$ for the peripheral case of $\bar{p} p$ 
         elastic scattering at energy of 541 GeV}
\end{figure}
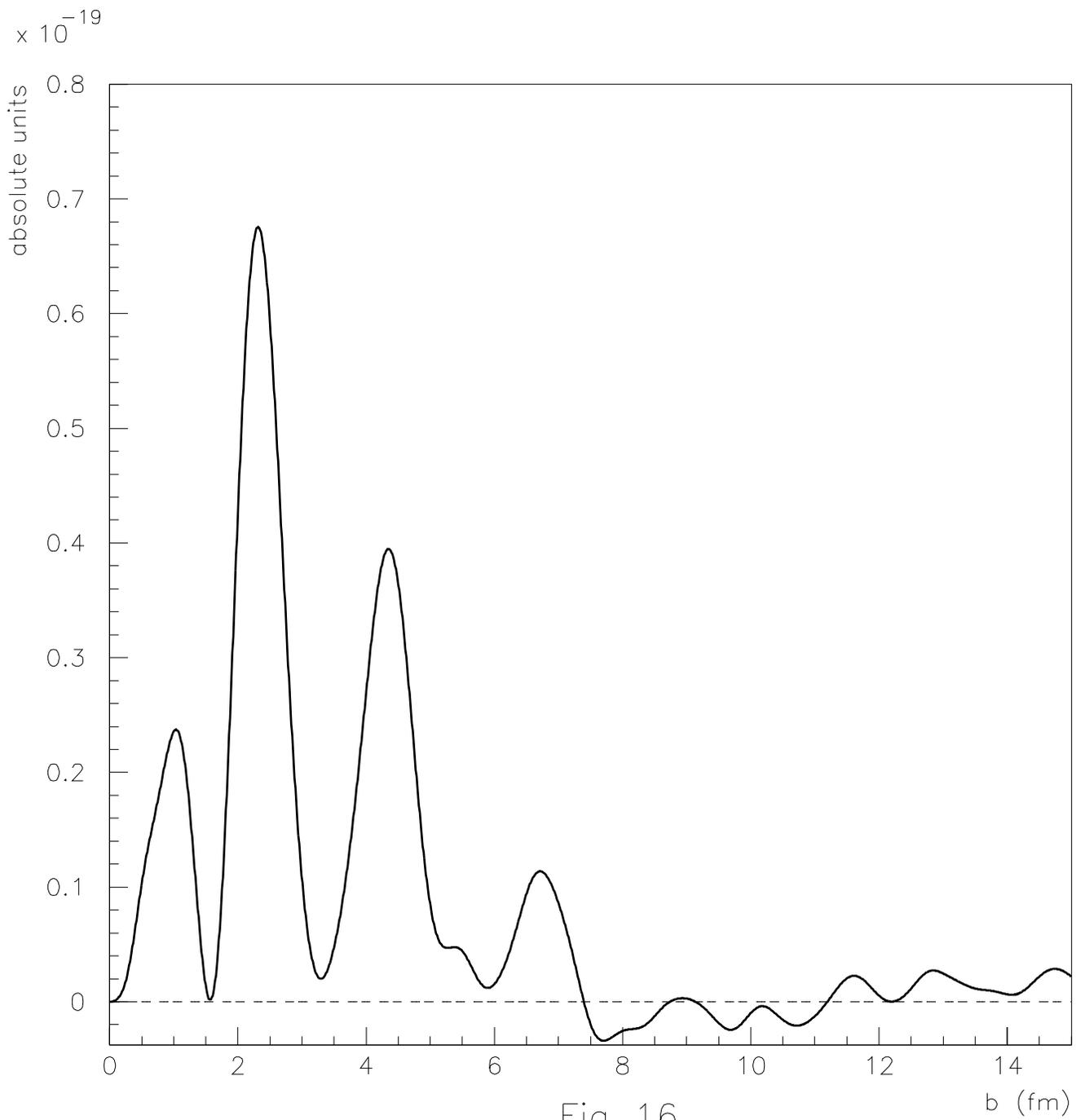
\begin{figure}	 
\caption{Correction function $K(s,b)$ calculated
         with the help of Eq. (\ref{cf1}) for the peripheral 
	 case of $\bar{p} p$ elastic scattering at energy of 541 GeV}
\end{figure}

\newpage

\begin{table}
\caption{Region of admissible 
values with typical statistical errors obtained by 
fitting the data of $pp$ elastic scattering at energy of
53 GeV and $\bar{p} p$ elastic scattering at energy of
541 GeV with the help of different formulas for
total elastic scattering amplitude}
\vspace*{1cm}
\begin{tabular}{ccccccc}

\hline \hline
 &  &  &  &  &  &  \\
data &ampl.&profile& $\sigma_{tot}$  &  $B$   & $\rho$ & $\chi^2/df$  \\
 &     &   & [mb] & [GeV$^{-2}$]&        &   \\
 &  &  &  &  &  & \\
\cline{1-7}

 &  &  &  &  &  & \\

$pp$ &(\ref{ta1})&peripheral&$42.65\div42.97$&$13.45\div 13.68$&$-0.12\div 0.08$ &
 252/201

                                       \\
  &  &           &($\pm0.12$)&($\pm0.05)$& ($\pm0.009$) &$\div$ 254/201 \\
 &  &  &  &  &  & \\                                       				       
    \cline{2-7}
 &  &  &  &  & \\
 &(\ref{ta1})& central&$42.65\div42.75$&$13.25\div 13.35$&$0.07\div0.08$ & 329/204     
                                        \\
53 GeV  &  &          & ($\pm0.23$)&($\div0.05$)&($\div0.009$)&$\div$339/204
  
                                        \\
&  &  &  &  &  & \\ 
 \cline{2-7} 
&  &  &  &  &  & \\					
 & W-Y &central&$42.38\pm0.15$&$12.87\pm0.14$&$0.077\pm0.009$ & 1.43
                       \\
 &  &  &  &  &  & \\ 		       
   \cline{1-7}					 
  &  &  &  &  &  & \\
$\bar{p} p$& (\ref{ta1})   &peripheral&$60.70\div 63.30$&$16.20\div16.55$&$0.11\div0.18$&233/213
                              \\                              
&    &          &($ \pm1.16 $)&$ (\pm0.09) $&$ (\pm0.022)$&$\div$238/213
                              \\
    &  &  &  &  &  & \\ 
  \cline{3-7}
       & &  &  &  &  &  \\
 
 &  (\ref{ta1})  &central&$62.20\div63.00$&$15.80\div16.00$&
 $0.08\div0.14$& 354/217
                                         \\
541 GeV   &   &         & ($\pm0.78$)& ($\pm0.05$)& ($\pm0.013)$& $\div$364/217
                                         \\  					 
&  &  &  &  &  & \\			      
\cline{2-7}
 &  &  &  &  &  & \\			       
& W-Y &central&$62.17\pm1.50$&$15.50\pm0.10$&$0.135\pm0.015$ & 1.1
                                        \\ 
 &  &  &  &  &  &  \\
\hline
\hline
\end{tabular} 
\end{table}

\begin{table}
\caption{Values of the total, elastic and inelastic
root-mean-squares calculated with the help of formulas 
(\ref{mv8}), (\ref{mv4}) and (\ref{mv13}) for 
$pp$ elastic scattering at energy of 53 GeV
$\bar{p} p$ elastic scattering at energy of 541 GeV}
\vspace*{1cm}
\begin{tabular}{ccccccc}
\hline
\hline

         &  &  & \multicolumn{3}{c}{} & \\
   data &profile   & $\sqrt{<b^2>_{tot}}$ & 
     \multicolumn{3}{c}{$\sqrt{<b^2>_{el}} \; \;$  }
    & $\sqrt{<b^2>_{inel}}$  \\      
    &  &  &    modulus & phase & total &   \\ 
          &  &[fm]  &[fm]  &[fm]  &[fm]  &[fm] \\ 
     &  &  &  &  &  & \\
	   \cline{1-7}
     &  &  &  &  &  & \\
      
 $pp$    &peripheral& 1.026$\div$1.033 & 0.68 &1.64$\div$1.67 
     &1.78$\div$1.80 &   0.76$\div$0.78\\ 
     &  &  &  &  &  &  \\   \cline{2-7}
     &  &  &  &  &  &  \\
  
  53 GeV      &central& 1.03 & 0.68 & $\sim$ 0. & 0.68 & 1.09   \\  
       &  &  &  &  &  &  \\ \cline{1-7}
     &  &  &  &  &  & \\
     
 $\bar{p} p$    &peripheral& 1.123$\div$1.135 & 0.76 &2.11$\div$2.22 &
   2.24$\div$2.35 &0.39$\div$0.49 \\ 
     &  &  &  &  &  &  \\   \cline{2-7}
     &  &  &  &  &  &  \\

 541 GeV    & central& 1.13 & 0.76 & $\sim$ 0. & 0.76 & 1.21       \\
     &  &  &  &  &  & \\
     \hline 
                \hline
   \end{tabular}
   
\end{table}

\end{document}